\documentclass[aps,pra,reprint,amsmath,amssymb,amsfonts,superscriptaddress]{revtex4-1}
\pdfoutput=1

\usepackage{graphicx}
\usepackage{hyperref}
\usepackage{mathtools}
\usepackage{siunitx}
\usepackage{esdiff}

\newcommand{\im}{\mathrm{i}}
\newcommand{\vect}[1]{\boldsymbol{\mathrm{#1}}}
\newcommand{\dif}{\mathrm{d}}
\newcommand{\e}[1]{e^{#1}}
\newcommand{\conj}{^\ast}
\newcommand{\inv}{^{-1}}
\newcommand{\dagg}{^\dagger}

\newcommand{\paren}[1]{\left(#1\right)}

\newcommand{\sbrak}[1]{\left[#1\right]}
\newcommand{\vbrak}[1]{\left\langle#1\right\rangle}

\newcommand{\ket}[1]{\lvert#1\rangle}

\newcommand{\Paren}[1]{\bigl(#1\bigr)}
\newcommand{\Sbrak}[1]{\bigl[#1\bigr]}

\newcommand{\Vbrak}[1]{\bigl\langle#1\bigr\rangle}
\newcommand{\Set}[1]{\bigl\{#1\bigr\}}

\DeclareMathOperator{\sinc}{sinc}
\DeclareMathOperator{\tr}{tr}

\newcommand{\sig}{_\text{s}}
\newcommand{\pump}{_\text{p}}

\begin{document}
\title{Nonlinear Quantum Behavior of Ultrashort-Pulse Optical Parametric Oscillators}

\author{Tatsuhiro~Onodera}
\thanks{These authors contributed equally to this work. Email:\\tatsuhiro.onodera@ntt-research.com\\edwin.ng@ntt-research.com}
\affiliation{Edward L.\ Ginzton Laboratory, Stanford University, Stanford, California 94305, USA}
\affiliation{Physics \& Informatics Laboratories, NTT Research, Inc., Sunnyvale, California 94085, USA}
\affiliation{School of Applied and Engineering Physics, Cornell University, Ithaca, New York 14853, USA}

\author{Edwin~Ng}
\thanks{These authors contributed equally to this work. Email:\\tatsuhiro.onodera@ntt-research.com\\edwin.ng@ntt-research.com}
\affiliation{Edward L.\ Ginzton Laboratory, Stanford University, Stanford, California 94305, USA}
\affiliation{Physics \& Informatics Laboratories, NTT Research, Inc., Sunnyvale, California 94085, USA}

\author{Chris~Gustin}
\thanks{These authors contributed equally to this work. Email:\\tatsuhiro.onodera@ntt-research.com\\edwin.ng@ntt-research.com}
\affiliation{Edward L.\ Ginzton Laboratory, Stanford University, Stanford, California 94305, USA}

\author{Niels~L\"orch}
\affiliation{Edward L.\ Ginzton Laboratory, Stanford University, Stanford, California 94305, USA}
\affiliation{Department of Physics, University of Basel, Klingelbergstrasse 82, 4056 Basel, Switzerland}

\author{Atsushi~Yamamura}
\affiliation{Edward L.\ Ginzton Laboratory, Stanford University, Stanford, California 94305, USA}

\author{Ryan~Hamerly}
\affiliation{Edward L.\ Ginzton Laboratory, Stanford University, Stanford, California 94305, USA}
\affiliation{Physics \& Informatics Laboratories, NTT Research, Inc., Sunnyvale, California 94085, USA}
\affiliation{Research Laboratory of Electronics, Massachusetts Institute of Technology, Cambridge, Massachusetts 02139, USA}

\author{Peter~L.~McMahon}
\affiliation{School of Applied and Engineering Physics, Cornell University, Ithaca, New York 14853, USA}

\author{Alireza~Marandi}
\affiliation{Department of Electrical Engineering, California Institute of Technology, Pasadena, California 91125, USA}

\author{Hideo~Mabuchi}
\affiliation{Edward L.\ Ginzton Laboratory, Stanford University, Stanford, California 94305, USA}

\date{\today}

\begin{abstract}
The quantum features of ultrashort-pulse optical parametric oscillators (OPOs) are investigated theoretically in the nonlinear regime near and above threshold. Viewing the pulsed OPO as a multimode open quantum system, we rigorously derive a general input-output model that features nonlinear coupling among many cavity (i.e., system) signal modes and a broadband single-pass (i.e., reservoir) pump field. Under appropriate assumptions, our model produces a Lindblad master equation with multimode nonlinear Lindblad operators describing two-photon dissipation and a multimode four-wave-mixing Hamiltonian describing a broadband, dispersive optical cascade, which we show is required to preserve causality. To simplify the multimode complexity of the model, we employ a supermode decomposition to perform numerical simulations in the regime where the pulsed supermodes experience strong single-photon nonlinearity. We find that the quantum nonlinear dynamics induces pump depletion as well as corrections to the below-threshold squeezing spectrum predicted by linearized models. We also observe the formation of non-Gaussian states with Wigner-function negativity and show that the multimode interactions with the pump, both dissipative and dispersive, can act as effective decoherence channels. Finally, we briefly discuss some experimental considerations for potentially observing such quantum nonlinear phenomena with ultrashort-pulse OPOs on nonlinear nanophotonic platforms.
\end{abstract}

\maketitle

\section{Introduction}
Ultrashort-pulse optical parametric oscillators (OPOs) have become established as an ideal testbed for the generation and manipulation of coherent nonlinear interactions among many optical frequency modes at once. In the classical domain, pulsed OPOs are used to generate frequency combs for applications in molecular spectroscopy and atomic clocks~\cite{Adler2009, Kippenberg, Spaun2016, Marandi2016}, and the strong temporal confinement of the field facilitates efficient nonlinear optics~\cite{Boyd2008}. In quantum experiments, they have been synchronously pumped below threshold to generate multimode squeezed light~\cite{Roslund2014, Gerke2015, Shelby1992}. Because their quantum states intrinsically reside in a multimode Hilbert space of high dimensionality, they are also being investigated as a resource for optical quantum information processing~\cite{Menicucci2006, Ferrini2014, Brecht2015, Cai2017, ra2020non}.

Many of the quantum features of pulsed OPOs are inherited from their single-mode continuous-wave (cw) counterparts, including squeezing~\cite{Wu1987}, non-Gaussian state generation~\cite{Reid1993}, and their applications to quantum information and communication~\cite{Braunstein2005, Gehring2015}. Quantum input-output theory, which describes open quantum system dynamics using master equations in Lindblad form~\cite{Gardiner1985a, Wiseman2010, Combes2017}, has been pivotal in elucidating the properties of cw OPOs. This framework makes numerical simulation tractable by reducing the OPO physics to the dynamics of a single internal mode interacting with a white-noise reservoir~\cite{Gardiner1992a, Wiseman1993}. As a result, a number of sophisticated techniques, from quantum measurement and feedback to quantum coherent control~\cite{Combes2017, Carmichael1993}, have been applied to the analysis of cw OPO dynamics and networks~\cite{Crisafulli2013, Iida2012}.

In this context, it is interesting to ask whether quantum input-output theory can be applied to pulsed OPOs as well. At first glance, such a theory appears intractable due to the large number of internal cavity modes (typically \num{e4} to \num{e5}). Nevertheless, such a construction was successfully demonstrated by Refs.~\cite{DeValcarcel2006, Patera2010, Navarrete-Benlloch2017} for synchronously pumped OPOs (SPOPOs). A key technique employed in these seminal papers was to recast the multimode input-output model into a low-dimensional supermode basis~\cite{Brecht2015, Wasilewski2006} over the signal resonances. This concise description enabled a detailed analysis~\cite{DeValcarcel2006, Patera2010, Navarrete-Benlloch2017} of the multimode Gaussian states produced in table-top SPOPOs, which, as experimentally demonstrated in Ref.~\cite{Roslund2014}, manifest as rich, highly entangled optical networks.

On the other hand, non-Gaussian states can arise in systems with nonlinear dynamics, provided sufficiently strong single-photon nonlinearities. For instance, cw OPOs in the deeply quantum regime have been theoretically predicted to produce Schr\"odinger cat states~\cite{Reid1993}, which can form the basis for schemes in quantum computation~\cite{Mirrahimi2014c, Gilchrist2004} and quantum-enhanced metrology~\cite{Gilchrist2004, Munro2002}. Though this regime has thus far only been accessible in ``exotic'' quantum systems such as in atom-cavity~\cite{Peyronel2012} or superconducting-circuit~\cite{Kirchmair2013} quantum electrodynamics, recent rapid advances in thin-film integrated nanophotonics suggest single-photon nonlinearities may soon also be accessible with all-optical (i.e., $\chi^{(2)}$ or $\chi^{(3)}$) nonlinearities, due to the strong spatial confinement of light into sub-wavelength nonlinear waveguides~\cite{lu2020toward, wang2018ultrahigh, park2021high}. Combined with the promising potential of these platforms to support advanced dispersion engineering (and hence strong temporal confinement as well)~\cite{jankowski2020ultrabroadband, jankowski2021dispersion}, ultrashort-pulse SPOPOs exhibiting few-photon nonlinear quantum dynamics appear to be within the realm of experimental possibility.

In this paper, we study theoretically the quantum behavior of ultrashort-pulse OPOs in this highly nonlinear regime. Our analysis is based on a rigorous, general quantum input-output model of an SPOPO, valid in both Gaussian and non-Gaussian regimes of operation. We model the nonlinear three-wave interaction between nonresonant pump and resonant signal modes using a nonlinear system-reservoir Hamiltonian and derive input-output relations from the Heisenberg equations of motion, while the internal dynamics of the SPOPO are captured with a time-convolutionless second-order Born-Markov master equation. In the process, we define ``band-limited'' quantum noise operators, which reduce to the noise operators in Ref.~\cite{Patera2010} under appropriate timescale limits. In the process, we thus clarify the requirements for a quantum input-output model---which is formulated in continuous time by construction---to be compatible with the pulsed nature of the system. The resulting model includes Lindblad operators representing nonlinear dissipation induced by pump depletion, but notably it also reveals the somewhat surprising existence of a nonlinear (quartic) dispersive Hamiltonian, which we show is necessary to preserve causality and maintain consistency with classical models in the mean-field limit.

We also show that the supermode technique of Refs.~\cite{DeValcarcel2006, Patera2010, Navarrete-Benlloch2017} can be applied to our model, in order to obtain an efficient description of the \emph{nonlinear} quantum dynamics. This approach enables us to perform numerical simulations in the supermode basis and observe a variety of nonlinear phenomena predicted by the model, such as pump depletion, corrections to the linearized squeezing spectrum, and the generation of non-Gaussian states. We provide estimates for the experimental parameters and regimes needed to observe these exotic quantum effects and compare them against the state of the art in nonlinear nanophotonics.

\section{Multimode input-output theory of pulsed OPOs} \label{sec:input-output-theory}
\begin{figure}[b!]
    \includegraphics[width=\linewidth]{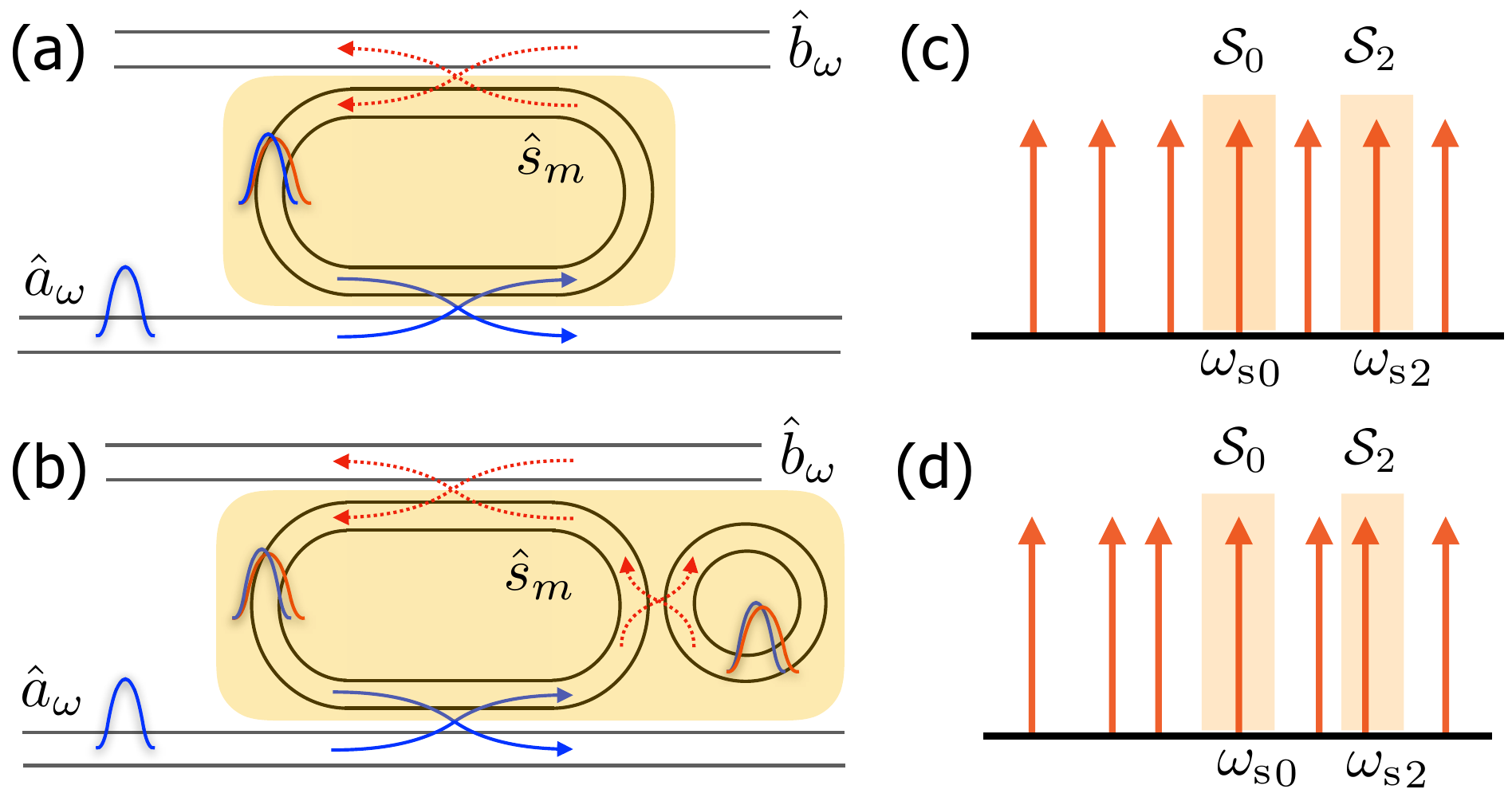}
    \caption{Schematics of various systems of pulsed OPOs and their cavity mode structure. (a) A single resonator (highlighted in yellow) is coupled on the bottom to a propagating pump field through a dichroic and on top to a propagating signal field through an outcoupler, resulting in (c) a set of uniformly spaced resonances. (b) Two coupled resonators form a single system (highlighted in yellow), resulting in (d) a nonuniform mode structure. In both cases, only the signal is resonant in the system, and the cavity medium (i.e., region in yellow) is taken to contain the $\chi^{(2)}$ nonlinearity (with appropriate dispersion compensation). The highlights in (c) and (d) indicate the bands $\mathcal S_m$ over which we define signal input-output operators (see main text).}
    \label{fig:schematic}
\end{figure}

The input-output formalism deals with systems coupled weakly to a reservoir, which we take to be a good characterization of a high-finesse pulsed OPO (the system) coupled to freely propagating optical fields (the reservoir)~\cite{Gardiner1985a, Wiseman2010}. The systems we consider consist of a broadband set of resonant modes in a ``signal'' band of frequencies and are schematically shown in Fig.~\ref{fig:schematic}. The system exhibits linear coupling to a corresponding signal band $\mathcal S$ in the free field and \emph{nonlinear} coupling to a second-harmonic ``pump'' band $\mathcal P$, also in the free field. Note that we use ``pump'' to refer to the reservoir with frequencies in the range of the second harmonic of the signal modes, even when there is no active pumping. Following the usual procedure for high-finesse optical systems, we assume the system (quasi)modes can be quantized independently of the reservoir, and we derive perturbatively the system dynamics subject to the effects of the reservoir in an input-output framework.

Let the cavity resonate a set of signal modes $\hat s_m$, with resonance frequencies ${\omega\sig}_m$. For the following, we work in an interaction frame rotating at these frequencies, generated by a Hamiltonian $\sum_m {\omega\sig}_m \hat{s}_m^{\dagger} \hat{s}_m$. We suppose the resonant signal modes are described by an electric displacement field operator $\hat{\vect D}\sig(\vect r, t)$ of the form
\begin{equation}
    \hat{\vect D}\sig(\vect r, t) = \im \sum_m \vect {D\sig}_m(\vect r) \, \hat s_m \e{-\im{\omega\sig}_m t} + \text{H.c.},
\end{equation}
where $\sbrak{\hat s_m, \hat s_n\dagg} = \delta_{mn}$, and ${\vect D\sig}_m(\vect r)$ are appropriately chosen mode functions, as prescribed by canonical quantization of the macroscopic Maxwell's equations~\cite{Drummond14, Quesada17}.

In the derivation to follow, we do not necessarily assume uniform cavity mode spacing, which is the example depicted in Fig.~\ref{fig:schematic}(c). This special case of uniform mode spacing is addressed in more detail in Sec.~\ref{sec:spopo}. In general, however, cavity resonances may not be uniformly spaced, either due to intracavity dispersion (especially in broadband cavities) or because the system may consist of multiple coupled cavities, as depicted in Fig.~\ref{fig:schematic}(b), resulting in the ``resonance splitting'' in Fig.~\ref{fig:schematic}(d). The following derivation can be applied to these nonuniform cases as well.

\subsection{Linear dissipation} \label{sec:linear-dissipation}
We begin by treating the linear coupling to the reservoir at the signal frequency band. Aside from outcoupling, this can also describe linear losses due to scattering or other intrinsic imperfections; while such effects are by nature spatially multimode, we take the usual assumption that, for each cavity mode, the various scattering channels can be combined into a single effective coupling to the reservoir, following Wigner-Weisskopf theory~\cite{Yamamoto1999}. 

We introduce reservoir modes $\hat b_\omega$, with $[\hat b_\omega, \hat b_{\omega'}\dagg] = 2\pi\delta(\omega-\omega')$ in a signal frequency range $\mathcal S$ of interest, and posit a minimal-coupling Hamiltonian in the interaction frame of the form
\begin{equation}
    \hat V_\text{lin}(t) \coloneqq \im \sum_m \int_{\mathcal S} \frac{\dif\omega}{2\pi} \; \sqrt{2\kappa_m(\omega)} \, \hat s_m \hat b_{\omega}\dagg \, \e{- \im ({\omega\sig}_m - \omega)t } + \text{H.c.} \label{eq:Vlin}
\end{equation}

To further develop this interaction in a way that suits a multimode cavity, we define a set of ``band-limited'' reservoir operators
\begin{align} \label{eq:sig-noise-operators}
    \hat b^{(m)}_t & \coloneqq \int_{\mathcal S_m} \!\!\frac{\dif\omega}{2\pi} \, \hat b_{\omega} \, \e{-\im (\omega-{\omega\sig}_m) t}, \quad\text{where}
    \\ \mathcal S_m & \coloneqq \Paren{\textstyle \frac12 \paren{{\omega\sig}_m + {\omega\sig}_{m-1}}, \frac12 \paren{{\omega\sig}_m + {\omega\sig}_{m+1}}},
\end{align}
as illustrated in Fig.~\ref{fig:schematic}(c,d). We can compute the evolution of these reservoir modes using their Heisenberg equations of motion. We parametrize this evolution by $t$, which we note is distinct in nature from the index $t'$ denoting the mode. This produces
\begin{align}
    &\frac{\dif}{\dif t}\hat b_{t'}^{(m)}(t) = -\im \Sbrak{\hat b_{t'}^{(m)}(t), \hat V_\text{lin}(t)} \label{eq:b_eom}\\
    &\quad{}= \sum_{n} \hat s_{n} e^{\im {\omega\sig}_m t'} e^{-\im {\omega\sig}_n t} \int_{\mathcal S_m} \frac{\dif\omega}{2\pi} \sqrt{2\kappa_{n}(\omega)} \e{+\im\omega(t-t')}. \nonumber
\end{align}
We now assume a \emph{Markov condition}, in which the bandwidths of both $\kappa_n(\omega)$ (i.e., the range of $\omega$ over which $\kappa_n(\omega)$ is sufficiently flat) and $\mathcal S_m$ (i.e., the quantity $\frac12\paren{{\omega\sig}_{m+1} - {\omega\sig}_{m-1}}$) are much larger than the bandwidth of any system dynamics. Following similar assumptions made in single-mode quantum optics, this Markov condition allows us to formulate a \emph{multimode} Markovian input-output theory for this interaction, and we will further develop and utilize this condition throughout this work. Under this Markov condition, the integral in \eqref{eq:b_eom} can effectively be replaced with $\sqrt{2\kappa_{n}(\omega_m)} \delta(t-t')$, and the result is simplified to
\begin{align}
    \frac{\dif}{\dif t}\hat b_{t'}^{(m)}(t) =  \sum_{n} \hat s_{n} e^{\im ({\omega\sig}_m - {\omega\sig}_{n}) t}\sqrt{2\kappa_{n}(\omega_m)} \delta(t-t').
\end{align}
From this, we see the reservoir mode only evolves discretely upon interacting with the system at $t'=t$, so we can isolate these two segments of its evolution as an input and output part: \begin{align}
    {\hat{b}_{t,\mathrm{in}}}^{(m)} &\coloneqq \lim_{t'\rightarrow-\infty} \hat{b}_{t}^{(m)}(t'), &
    {\hat{b}_{t,\mathrm{out}}}^{(m)} &\coloneqq \lim_{t'\rightarrow+\infty} \hat{b}_t^{(m)}(t').
\end{align}
which are related via the input-output relationship
\begin{align}
    \hat{b}_{t,\mathrm{out}}^{(m)}
    &= \hat{b}_{t,\mathrm{in}}^{(m)} + \sum_{n} \sqrt{2\kappa_n(\omega_m)} \hat{s}_n(t) e^{\im({\omega \sig}_{m} -{\omega \sig}_n)t} \nonumber\\
    & \approx \hat{b}_{t,\mathrm{in}}^{(m)} + \sqrt{2 \kappa_m}\hat{s}_m(t),\label{eq:inoutlin}
\end{align}
where $\kappa_m \coloneqq \kappa_m({\omega\sig}_m)$ and we have made a rotating-wave approximation in the second line, which holds as long as the Markov condition is satisfied. In this case, $\kappa_m$ is the field amplitude decay rate for the $m$th signal mode.

Next, we derive the evolution of the system signal modes due to the coupling $\hat V_\text{lin}$, in the form of a Lindblad master equation, starting from a second-order time-convolutionless Born-Markov approximation~\cite{Breuer2002}. Let $\hat\rho$ denote the system state, $\tr_{\bar{\mathcal S}}$ denote a partial trace over the reservoir, and $\hat\rho_{\bar{\mathcal S}}$ denote the density matrix of the reservoir in the absence of interaction with the system, which we take to correspond to the vacuum. The contribution to $\dif\hat\rho/\dif t$ due to $\hat V_\mathrm{lin}$ is given by
\begin{equation} \label{eq:me1}
    \mathcal L_{\rm lin} \hat\rho  = -\int_0^\infty \!\!\dif\tau \, \tr_{\bar{\mathcal S}} \!\sbrak{\hat V_\mathrm{lin}(t), \Sbrak{\hat V_\mathrm{lin}(t-\tau), \hat\rho(t)\hat\rho_{\bar{\mathcal S}}}}.
\end{equation}
In evaluating \eqref{eq:me1}, we neglect Lamb shifts (i.e., assume they can be absorbed into renormalized energy levels ${\omega\sig}_m$), and we make a secular (post-trace rotating-wave) approximation by dropping any terms in \eqref{eq:me1} that oscillate as $\e{-\im ({\omega\sig}_n-{\omega\sig}_m)t}$ for $m \neq n$, which is also justified by the Markov condition. The result is the standard dissipative contribution to the Lindblad master equation
\begin{subequations} \label{eq:me-lin}
\begin{equation}
    \mathcal L_\mathrm{lin}\hat\rho = \sum_m \mathcal D\Sbrak{\hat L^{(m)}_\text{lin}} \hat\rho,
\end{equation}
where $\mathcal D[\hat L]\hat\rho \coloneqq \hat L \hat\rho \hat L\dagg - \frac 1 2 \Set{\hat L\dagg\hat L, \hat\rho}$~\footnote{The anticommutator is $\Set{\hat A, \hat B} \coloneqq \hat A\hat B + \hat B \hat A$.}, and we have identified a set of linear Lindblad operators
\begin{equation}
    \hat L^{(m)}_\text{lin} \coloneqq \sqrt{2\kappa_m} \hat s_m
    \label{eq:Llin}
\end{equation}
\end{subequations}
representing dissipation into each frequency bin.

\subsection{Nonlinear parametric interactions}
\label{sec:io-theory-nl}

We now turn to the treatment of the nonlinear $\chi^{(2)}$ interaction between the cavity signal modes and the nonresonant pump field. We suppose the pump is described as a spectrally continuous field, with an electric displacement field operator $\hat{\vect D}\pump$ of the form
\begin{equation}
    \hat{\vect D}\pump(\vect r, t) = \im \int_{\mathcal P} \frac{\dif\omega}{2\pi} \, \vect {D\pump}_\omega(\vect r) \, \hat a_\omega \e{-\im{\omega} t} + \text{H.c.}
\end{equation}
in the interaction frame, where the continuum pump reservoir modes $\hat a_\omega$ obey $\Sbrak{\hat a_\omega, \hat a_{\omega'}\dagg} = 2\pi\delta(\omega-\omega')$ and ${\vect D\pump}_\omega(\vect r)$ are appropriately chosen continuum mode profile functions, as prescribed by canonical quantization of the macroscopic Maxwell's equations~\cite{Drummond14,Quesada17}. The frequency range $\mathcal P$ for this integral (i.e., the pump band) should be sufficiently confined so as to not overlap with the signal reservoir band $\mathcal S$. Note that in this subsection, we do not yet consider an active coherent drive on the pump field (i.e., we do not yet ``pump'' the OPO), but in Sec.~\ref{sec:spopo} we show how such a drive can be easily handled once we have derived the appropriate interaction with the pump reservoir.

We take the macroscopic nonlinear $\chi^{(2)}$ Hamiltonian in the interaction frame to be~\cite{Drummond14,Quesada17}
\begin{align}
    \label{eq:Vnl}
    &\hat V_\text{nl}(t) \coloneqq  \int \dif^3 \vect r \sum_{i,j,k}\, \eta^{(2)}_{ijk}(\vect r) {\hat D\pump}^i(\vect r) \hat D^j\sig(\vect r) \hat D^k\sig(\vect r) \\
    &\quad{}= \im \int_{\mathcal P} \frac{\dif\omega}{2\pi} \, \hat a_{\omega}\dagg \sum_{m,n} f_{mn}(\omega) \, \hat s_m \hat s_n \e{-\im({\omega\sig}_m + {\omega \sig}_n-\omega)t} + \text{H.c.}, \nonumber
\end{align}
where $\eta^{(2)}_{ijk}(\vect r)$ is the second-order inverse susceptibility tensor~\cite{Drummond14} (assumed to be frequency independent) and the coupling strength of the three-wave interaction is
\begin{equation} \label{eq:coupling-fmn}
    f_{mn}(\omega) \coloneqq \int \dif^3 \vect r \, \sum_{i,j,k} \eta^{(2)}_{ijk}(\vect r) \Paren{{{D^i\pump}\conj_\omega}{D^j\sig}_{\!m}{D^k\sig}_{\!n}}(\vect r),
\end{equation}
which we can take to be real-valued without loss of generality when ${\vect D\pump}_\omega(\vect r)$ corresponds to normal modes of the system. That is, linear loss of these modes have been phenomenologically treated according to the preceding subsection, as opposed to direct quantization of lossy (quasinormal) modes~\cite{Franke2019May}. In Sec.~\ref{sec:phase-matching}, we consider some concrete forms for $f_{mn}(\omega)$ in SPOPOs based on phase-matching considerations.

We follow a similar procedure as in the linear dissipation case in order to define input-output pump operators. The choice of partitions $\mathcal P_q$ of the pump frequencies is somewhat arbitrary; if the pump is a frequency comb, for example, one natural choice is to use the pump comb lines. In any case, let us denote such pump frequencies of interest by ${\omega\pump}_q$. We now define pump operators
\begin{align}
    \label{eq:pump-res-definition}
    \hat a_t^{(q)} &\coloneqq \int_{\mathcal P_q} \frac{\dif\omega}{2\pi} \, \hat a_{\omega} \e{-\im(\omega-{\omega\pump}_q)t}, \quad\text{where}
    \\ \mathcal P_q &\coloneqq \Paren{\textstyle \frac12 ({\omega\pump}_q+{\omega\pump}_{q-1}), \frac12 ({\omega\pump}_q+{\omega\pump}_{q+1})}.
\end{align}

To derive the evolution of these reservoir modes $\hat{a}_{t'}^{(q)}$, we again calculate
\begin{align}
    &\diff{}{t}\hat a_{t'}^{(q)}(t) = -\im \Sbrak{\hat a_{t'}^{(q)}(t), \hat V_\text{nl}(t)} \label{eq:a_eom}\\
    &= \sum_{m,n} \hat s_m \hat s_n \e{-\im({\omega\sig}_m + {\omega \sig}_n)t} \e{\im{\omega\pump}_q t'} \!\! \int_{\mathcal P_q} \frac{\dif\omega}{2\pi} f_{mn}(\omega) \e{+\im\omega(t-t')}. \nonumber
\end{align}
As in the case for linear dissipation, we impose the Markov condition that the bandwidths of the signal-pump coupling $f_{mn}(\omega)$ as well as the pump bands $\mathcal P_q$ are much larger than any system dynamical rate. Under this Markov condition, we can effectively replace the integral by $f_{mn}({\omega\pump}_q)\delta(t-t')$, and the result is
\begin{equation} \label{eq:a_eom_2}
    \diff{}{t}\hat a_{t'}^{(q)}(t) = \sum_{m,n} f_{mn}^{(q)} \hat s_m \hat s_n \e{-\im({\omega\sig}_m + {\omega \sig}_n-{\omega\pump}_q)t}\delta(t-t'),
\end{equation}
where we introduce the simplified notation
\begin{equation}
    f_{mn}^{(q)} \coloneqq f_{mn}({\omega\pump}_q).
\end{equation}
In contrast to the case of linear dissipation, the Markov condition on $f_{mn}(\omega)$ needs to be treated with some care. We show in Appendix~\ref{app:nl_tensor}, in the context of the Born-Markov master equation, that the bandwidth of $f_{mn}(\omega)$ is intimately linked to the memory time of the pump reservoir, which in turn is related to the roundtrip time of the signal cavity, consistent with the requirement for the bandwidths of $\mathcal P_q$ or $\mathcal S_m$ to be sufficiently large; we also provide a brief discussion of these timescale considerations for SPOPOs in Sec.~\ref{sec:discussion-approx}.

Again, this $\delta$-function interaction with the system in \eqref{eq:a_eom_2} produces an input-output relationship
\begin{subequations}\label{eq:nl_inout}
\begin{align}
    \hat{a}^{(q)}_{t, \rm out} = \hat{a}^{(q)}_{t, \rm in} + \sum_{m,n}f_{mn}^{(q)}\hat{s}_m \hat{s}_n e^{-i({\omega\sig}_m + {\omega \sig}_n -{\omega \pump}_{q})t},
\end{align}
where
\begin{align}
    {\hat{a}_{t,\mathrm{in}}}^{(q)} &\coloneqq \lim_{t'\rightarrow-\infty} \hat{a}_{t}^{(q)}(t'), &
    {\hat{a}_{t,\mathrm{out}}}^{(q)} &\coloneqq \lim_{t'\rightarrow+\infty} \hat{a}_t^{(q)}(t').
\end{align}
\end{subequations}
In the following section, we show that the rotating terms in \eqref{eq:nl_inout} can be eliminated for a synchronously pumped OPO with a natural choice for the pump frequencies ${\omega\pump}_q$. In general, however, if ${\omega\pump}_q$ are arbitrarily picked relative to the signal frequencies ${\omega\sig}_m$, one may not be able to apply a rotating-wave approximation to simplify this input-output relation further.

We next turn to deriving the master-equation model for the evolution of the signal modes subject to this nonlinear interaction with the pump reservoir. We again employ a second-order time-convolutionless Born-Markov approximation~\cite{Breuer2002}. Let $\hat\rho$ denote the system state, $\tr_{\bar{\mathcal P}}$ denote a partial trace over the pump reservoir, and $\hat\rho_{\bar{\mathcal P}}$ denote the density matrix of the pump reservoir in the absence of interaction with the system, which we take (for now) to be vacuum. The contribution to $\dif\hat\rho/\dif t$ by $\hat V_\text{nl}$ is
\begin{equation}
    \mathcal L_\mathrm{nl} \hat{\rho} = -\int_0^{\infty} \!\dif\tau \tr_{\bar{\mathcal P}} \sbrak{\hat{V}_{\rm nl}(t), \Sbrak{\hat{V}_\mathrm{nl}(t-\tau),\hat{\rho}(t)\hat{\rho}_{\bar{\mathcal P}}}}.
\label{eq:BM-master-eqn}
\end{equation}
Note that as the signal frequency band and pump frequency band reservoirs are independent (i.e., nonoverlapping), the Born-Markov approximation allows the influence of these reservoirs to be derived independently of one another.
Evaluating the partial trace, we find
\begin{align}\label{eq:fullnl}
    &\mathcal L_{\rm nl} \hat{\rho} = \sum_{m',n'}\sum_{m,n} e^{\im({\omega\sig}_{m'} + {\omega \sig}_{n'})t} e^{-\im({\omega\sig}_{m} + {\omega \sig}_{n})t} \\
    &\qquad\qquad{}\times \xi_{mnm'n'} \Sbrak{\hat{s}_{m}\hat{s}_{n}\hat{\rho}(t),\hat{s}^{\dagger}_{m'} \hat{s}^{\dagger}_{n'}} + \text{H.c.}, \nonumber
\end{align}
with an interaction tensor
\begin{align}
    \xi_{mnm'n'} \coloneqq \int_0^\infty \dif \tau\, h_{mnm'n'}(\tau),
    \label{eq:xi-h}
\end{align}
where
\begin{align} \label{eq:memory-function}
     h_{mnm'n'}(\tau) \coloneqq \int_0^{\infty} \frac{\dif \omega}{2\pi}\, &f_{m'n'}(\omega) f_{mn}(\omega) \\
     &\quad{}\times\e{\im({\omega\sig}_{m} + {\omega \sig}_{n}-\omega)\tau}\nonumber
\end{align}
are nonlinear memory functions associated with the system's nonlinear coupling to the pump reservoir, which correspond to the temporal decay of correlations between the system and the pump reservoir induced by the nonlinear interaction. For the Born-Markov approximation to hold in \eqref{eq:BM-master-eqn}, $h_{mnm'n'}(\tau)$ must vanish for $\tau$ larger than any system interaction timescale (i.e., the memory time should be short). Using additional assumptions in Sec.~\ref{sec:phase-matching} (with details in Appendix~\ref{app:nl_tensor}), we explicitly evaluate \eqref{eq:memory-function} and show that the memory function indeed satisfies this requirement for a broad class of ultrashort-pulse OPOs.

Using the Sokhotski-Plemelj theorem from complex analysis and assuming $f_{mn}(\omega)$ to be continuous,
the coupling constants can also be written in terms of their real and imaginary parts such that $\xi = \gamma + \im\chi$, where
\begin{subequations}\label{eq:nlcouplings} \begin{align}
    \gamma_{mnm'n'} &\coloneqq \frac12 f_{m'n'}({\omega\sig}_{m}+{\omega\sig}_{n}) f_{mn}({\omega\sig}_{m}+{\omega\sig}_{n}) \\
    \chi_{mnm'n'} &\coloneqq \text{P}\!\int_0^{\infty} \frac{\dif \omega}{2\pi} \frac{f_{m'n'}(\omega)f_{mn}(\omega)}{{\omega\sig}_{m} + {\omega\sig}_{n}-\omega}, \label{eq:chi_cauchy_full} 
\end{align} \end{subequations}
where $\text{P}$ denotes the Cauchy principal value. As we will see later, under assumptions appropriate to synchronously pumped OPOs, $\gamma$ physically contributes to dissipative evolution under \eqref{eq:BM-master-eqn} while $\chi$ physically contributes to coherent evolution under \eqref{eq:BM-master-eqn}. However, we note that, as we show in Sec.~\ref{sec:phase-matching}, it can be easier in practice to directly compute $\xi$ using \eqref{eq:xi-h} and \eqref{eq:memory-function}, rather than the integrals in \eqref{eq:nlcouplings}.

Without additional assumptions, \eqref{eq:fullnl} is the most general quantum master-equation model for the nonlinear evolution of the signal modes. In general, it is a time-dependent master equation with nonlinear, \emph{non-Lindblad} dissipative terms (corresponding to $\gamma$), as well as an additional nonlinear effective Hamiltonian (corresponding to $\chi$). In Sec.~\ref{sec:spopo}, we show that a secular approximation, retaining only terms in the sum where ${\omega\sig}_m + {\omega\sig}_n \approx {\omega\sig}_{m'} + {\omega\sig}_{n'}$, can be made in \eqref{eq:fullnl} for the case of a synchronously pumped OPO, in which case we can further simplify \eqref{eq:fullnl} by casting it in Lindblad form. Then, as we show in Sec.~\ref{sec:phase-matching}, we can explicitly calculate the coupling constants in \eqref{eq:nlcouplings} once given a form for the phase-matching function $f_{mn}(\omega)$.

\section{The quasi-degenerate synchronously pumped OPO} \label{sec:spopo}
An experimentally relevant special case of the above theory applies to the quasi-degenerate SPOPO~\cite{Roslund2014, Gerke2015, McMahon2016}, in which
\begin{itemize}
    \item the system consists of a cavity resonating a uniform comb of signal modes with free spectral range $\Omega$ (i.e., any mode dispersion due to the nonlinear medium is compensated elsewhere in the cavity), as illustrated in Fig.~\ref{fig:schematic}(c);
    \item the system is pumped by a classical frequency comb (e.g., as produced by a mode-locked laser), with comb spacing (i.e., pulse repetition rate) equal to $\Omega$ (\emph{synchronous pumping});
    \item the phase matching is chosen such that maximal nonlinear coupling strength occurs between the center signal mode at frequency $\omega_0$ and the center pump line at frequency $2\omega_0$, but there is still sufficient phase matching off-center to facilitate nondegenerate interactions within the system optical bandwidth (\emph{quasi-degenerate}).
\end{itemize}
For such a system, it is convenient to enumerate the cavity signal modes as ${\omega\sig}_m = \omega_0 + m \Omega$ and the pump frequencies as ${\omega\pump}_q = 2\omega_0 + q \Omega$. In this case, the size of the frequency bands $\mathcal S_m$ and $\mathcal P_q$ are all given by $\Omega$. Thus, for SPOPOs, the Markov condition imposes the requirement that $\Omega$ be larger than all system dynamical rates (see Sec.~\ref{sec:discussion-approx} for addition discussion).

Assuming the Markov condition holds, the uniformity of the frequency spacings also enables a secular (rotating-wave) approximation to allow only near-resonant nonlinear interactions, such that the only contributions to the sum in \eqref{eq:nl_inout} obey $m+n=q$, and the only contributions to the sum in \eqref{eq:fullnl} obey $m+n=m'+n'$. Thus, the input-output relations \eqref{eq:nl_inout} for the pump reservoir can be rewritten for an SPOPO in the further simplified form
\begin{equation}
    \hat{a}^{(q)}_{t, \rm out} = \hat{a}^{(q)}_{t, \rm in} + \sum_{m+n=q}f_{mn}^{(q)}\hat{s}_m \hat{s}_n. 
    \label{eq:io_pump}
\end{equation}
In addition, the system dynamics under the Born-Markov master equation \eqref{eq:fullnl} can be simplified to
\begin{equation}\label{eq:secular_me}
    \mathcal L_\text{nl}\hat\rho = \sum_q \sum_{m,m'} \xi^{(q)}_{mm'} \sbrak{\hat s_m \hat s_{q-m} \hat\rho, \hat s_{m'}\dagg \hat s_{q-m'}\dagg} + \text{H.c.},
\end{equation}
where we have used $m+n = m'+n'=q$ in the inner sum to eliminate $n$ and $n'$. We have also defined $\xi^{(q)}_{mm'} \coloneqq \xi_{m,q-m,m',q-m'}$, so we can write
\begin{align}
\xi^{(q)}_{mm'} = \gamma^{(q)}_{mm'} + \im \chi^{(q)}_{mm'},
\label{eq:xi_mm'}
\end{align}
where $\gamma^{(q)}_{mm'}$ and $\chi^{(q)}_{mm'}$ are both real and symmetric over the $m$ and $m'$ indices. They are given by
\begin{subequations}
\begin{align}
    \gamma^{(q)}_{mm'} &\coloneqq \frac12 f^{(q)}_{m',q-m'} f^{(q)}_{m,q-m} \label{eq:gamma_mm'}\\
    \chi^{(q)}_{mm'} &\coloneqq \text{P}\! \int_0^\infty \frac{\dif\omega}{2\pi} \frac{f_{m',q-m'}(\omega) f_{m,q-m}(\omega)}{{\omega\pump}_q - \omega}
    \label{eq:chi_cauchy_secular} 
\end{align}
\end{subequations}
Using these expressions, we can now explicitly evaluate the commutators in \eqref{eq:secular_me} and obtain the nonlinear contribution to the Lindblad master equation
\begin{subequations} \label{eq:me-nonlin-spopo}
\begin{equation}
    \mathcal L_\mathrm{nl}\hat \rho = \sum_q \mathcal D\Sbrak{\hat L^{(q)}_\text{nl}} \hat\rho - \im \Sbrak{\hat H_\mathrm{nl}, \hat\rho},
\end{equation}
where the dissipative part of the evolution is contributed by $\gamma^{(q)}$ and consists of a multimode set of two-photon loss channels
\begin{equation} \label{eq:Lnl}
    \hat L_\text{nl}^{(q)} \coloneqq \sum_m f^{(q)}_{m,q-m} \hat s_m \hat s_{q-m} = \sum_{m+n=q} f^{(q)}_{mn} \hat s_m \hat s_n,
\end{equation}
while the coherent part is contributed by $\chi^{(q)}$ and takes the form of a four-wave-mixing dispersive optical cascade
\begin{equation}\label{eq:kerr}
    \hat{H}_\text{nl} = \sum_q \sum_{m,m'} \chi^{(q)}_{mm'} \hat s_{m'}\dagg \hat s_{q-m'}\dagg \hat s_m \hat s_{q-m}.
\end{equation}
\end{subequations}
As we show in Appendix~\ref{app:kk}, the existence of a cascaded nonlinear term in the system Hamiltonian is generally required to preserve causality in the presence of nonlinear dissipation, by virtue of a Kramers-Kronig-like relationship between the real and imaginary parts of the tensor $\xi$. While single-mode (cw) OPOs have the option of avoiding this effect by employing perfect phase matching, the same is not true in general for SPOPOs, where the large number of signal modes means that many elements of $\chi^{(q)}_{mm'}$ are nonzero due to dispersion.

Having derived a suitable input-output model for the interaction between the signal and pump, we also straightforwardly obtain a rigorous model for the important special case of an SPOPO with an \emph{active pump}. Suppose we drive the SPOPO with coherent pump amplitude $\alpha_q$ at frequency ${\omega\pump}_q$. To model this drive, one simply needs to displace $\hat a^{(q)}_{t,\text{in}}$ by $\alpha^{(q)}$ (so that $\Vbrak{\hat a^{(q)}_{t,\text{in}}} = \alpha^{(q)}$), which also adds a new system Hamiltonian $\im\sum_q \alpha^{(q)*} \hat L_\text{nl}^{(q)} + \text{H.c}$. However, in this paper, it is more convenient to keep the input in the vacuum (so that $\Vbrak{\hat a^{(q)}_{t,\text{in}}} = 0$) and instead apply an equivalent formal procedure~\cite{Combes2017} where we displace the Lindblad operators
\begin{subequations} \label{eq:pumped-spopo}
\begin{equation} \label{eq:nl-Lindblads-displaced}
    \hat L_\text{nl}^{(q)} \mapsto \sum_{m+n=q} f^{(q)}_{mn} \hat s_m \hat s_n + \alpha^{(q)},
\end{equation}
and add a system Hamiltonian
\begin{equation} \label{eq:H-pumped-spopo}
    \hat H_\text{pump} = \frac{\im}{2} \! \sum_{m,n} {\alpha^{(m+n)*}} f_{mn}^{(m+n)} \hat s_m \hat s_n + \text{H.c}.
\end{equation}
\end{subequations}
It is worth noting these two conventions for the input-output fields lead to the same Lindblad master equation \eqref{eq:me-nonlin-spopo}~\cite{Wiseman2010}. The pumped Hamiltonian \eqref{eq:H-pumped-spopo} describes multimode squeezing (i.e., a broadband version of the usual quadratic Hamiltonian for an OPO below threshold), and it is in agreement with prior derivations through other means, such as by assuming a resonant but adiabatically eliminated multimode pump~\cite{Patera2010}.

Finally, it is also often useful to analyze the experimentally relevant situation where the SPOPO possesses some slight nonuniformity in its signal resonances so that the bare frequencies of the modes are instead ${\omega\sig}_m = \omega_0 + m\Omega + \delta_m$, where $|\delta_m| \ll \Omega$ for the Markov condition to hold. First, we note that, throughout Sec.~\ref{sec:io-theory-nl}, we have performed all our calculations in a frame rotating at the bare frequencies of each signal mode so the results of Sec.~\ref{sec:io-theory-nl} need not change in principle. However, in trying to apply the secular approximation to arrive at \eqref{eq:secular_me} for the SPOPO, we cannot consistently define a unique set of ${\omega\pump}_q$ such that ${\omega\pump}_q = {\omega\sig}_m + {\omega\sig}_n = (m+n)\Omega + \delta_m + \delta_n$ for all $m + n = q$, due to the detunings being potentially unique for each signal mode. Thus even with a post-trace rotating-wave approximation, the master equation cannot be put into Lindblad form, and the system interactions involve the full rank-four tensor $\xi_{mnm'n'}$, which contains vastly more elements than the simplified $\xi^{(q)}_{mm'}$. To remedy this, we note that the derivation of Sec.~\ref{sec:input-output-theory} can instead be done starting from an interaction frame rotating at the \emph{nominal} mode frequencies $\omega_0 + m \Omega$ (i.e., without the perturbations $\delta_m$). In doing so, it is necessary to add a new ``detuning'' system Hamiltonian
\begin{equation} \label{eq:h0}
    \hat H_\mathrm{detuning} = \sum_m \delta_m \hat s_m^\dagger \hat s_m.
\end{equation}
Under the Markov condition, the effect of this detuning system Hamiltonian can be neglected while deriving the system-reservoir interactions in Sec.~\ref{sec:io-theory-nl}. In this paper, we only consider the case of $\delta_m = 0$ for our numerical simulations (so the bare and nominal mode frequency interaction frames coincide), but we retain the possibility of such inhomogeneities in our model for the sake of generality and to facilitate studying the robustness of our model to experimental imperfections. We reiterate that if the deviations $\delta_m$ are not small compared to the nominal mode spacing $\Omega$, however, the master equation and input-output model must be derived in the interaction frame of the bare frequencies according to Sec.~\ref{sec:input-output-theory}.

\subsection{Input-output theory and quantum stochastic differential equations}
Because the secular approximation allows us to describe the dynamics of the SPOPO using a master equation in standard Lindblad form, we can now formulate a standard input-output quantum model for the SPOPO, encompassing the multimode and nonlinear nature of its system-reservoir interactions.

Revisiting \eqref{eq:inoutlin} and \eqref{eq:Llin} for the linear dissipation and \eqref{eq:io_pump} and \eqref{eq:Lnl} for the nonlinear dissipation, we see that the input-ouput relations can be written in terms of the Lindblad operators \eqref{eq:Llin} and \eqref{eq:Lnl} (and \eqref{eq:nl-Lindblads-displaced} for active pumping) as
\begin{subequations}
\begin{align}
    \hat b_{t,\mathrm{out}}^{(m)} &= \hat b_{t,\mathrm{in}}^{(m)} + \hat L_\text{lin}^{(m)} \\
    \hat a_{t,\mathrm{out}}^{(q)} &= \hat a_{t,\mathrm{in}}^{(q)} + \hat L_\text{nl}^{(q)}.
\end{align}
\end{subequations}
Since the internal dynamics of the system are also governed by a master equation in Lindblad form using the same Lindblad operators, we can summarize all the dynamics of the SPOPO via an ``input-output model'' with total effective Hamiltonian
\begin{align}
    \label{eq:io_Hamiltonian}
    &\hat H_\text{spopo} = \hat{H}_\mathrm{detuning} + \hat{H}_\mathrm{nl} + \hat{H}_\mathrm{pump} \\
    &\qquad{}+ \paren{\im \sum_{m} \hat b^{(m)\dagger}_t \hat L^{(m)}_\text{lin} +\im \sum_{q} \hat a^{(q)\dagger}_t \hat L^{(q)}_\text{nl} + \text{H.c.}}.\nonumber
\end{align}
In this formulation of the model, it is crucial that the reservoir modes obey a \emph{white-noise approximation} that $\Sbrak{\hat b^{(m)}_t, \hat b^{(m')\dagger}_{t'}} \approx \delta(t-t')$, which is afforded by the Markov condition. Specifically, this commutator is given by
\begin{align}
    \Sbrak{\hat b^{(m)}_t, \hat b^{(m')\dagger}_{t'}} &= \delta_{mm'}\int_{-\Omega/2}^{\Omega/2} \frac{\dif\omega}{2\pi} \, \e{-\im \omega (t-t')},
    \label{eq:b_com}
\end{align}
and similarly for the pump reservoir operators $\hat a^{(q)}_t$. We therefore see that as long as the Markov condition holds, this integral can effectively be approximated by $\delta(t-t')$, hence justifying an interpretation of these reservoir operators as \emph{quantum white-noise operators}~\cite{Gardiner1985a}. This abstracted model of the SPOPO enables us to readily integrate SPOPOs into physical systems involving other quantum input-output devices through the use of the SLH formalism~\cite{Combes2017}. This system-level approach to quantum optics facilitates the construction of complex quantum networks and provides a powerful framework for deploying techniques such as dissipation engineering or quantum control.

Furthermore, the input-output theory for the SPOPO also allows us to formally derive quantum stochastic differential equations (QSDEs), or Heisenberg-Langevin equations, describing system dynamics subject to a quantum white-noise bath. In It\^o form, the QSDEs for the SPOPO are given by
\begin{align} \label{eq:eoms_bare_freq}
    \frac{\dif\hat s_m}{\dif t} = & -(\kappa_m + \im\delta_m)\hat s_m - 2\sum_{q}f^{(q)}_{m, q-m}\alpha^{(q)}\hat s\dagg_{q-m} \nonumber\\
    & -2\sum_q \sum_n \paren{\gamma^{(q)}_{nm} + \im  \chi^{(q)}_{nm} }  \hat s\dagg_{q-m} \hat s_n \hat s_{q-n} \nonumber \\
    & - \sum_q \sqrt{2\kappa_m} \hat b_{t,\text{in}}^{(m)} - 2\sum_{q}f^{(q)}_{m, q-m}\hat s\dagg_{q-m}\hat a_{t,\text{in}}^{(q)}. 
\end{align}
Provided all the approximations leading up to the quantum input-output model \eqref{eq:io_Hamiltonian} hold, these QSDEs concisely summarize all the quantum dynamics that can occur in an SPOPO. In addition to multimode squeezing in the linearized regime (as extensively studied in Ref.~\cite{Patera2010}), there is also a rich set of multimode nonlinear interactions above threshold as well, with both dissipative and dispersive contributions to the quantum dynamics.

\section{Phase matching} \label{sec:phase-matching}
\begin{figure*}[t!]
    \includegraphics[width=0.85\textwidth]{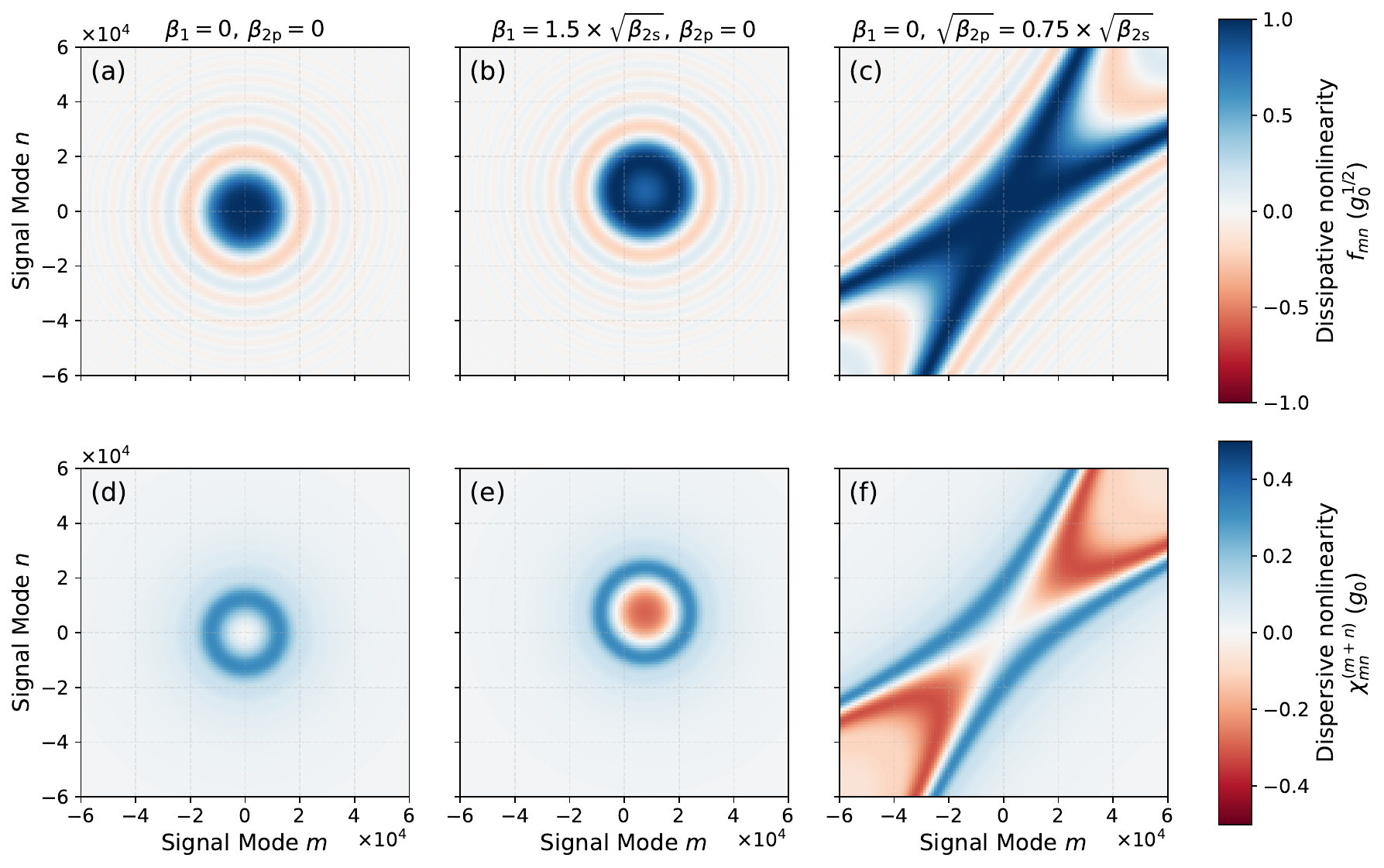}
    \caption{Structure of frequency-mode coupling coefficients for an SPOPO with $g_{mn} \approx g_0$, for various material dispersion parameters chosen according to the second-order expansion \eqref{eq:coupling-Phimn-expanded}. (a--c) Coupling coefficients $f_{mn}$ for the dissipative nonlinear interaction, given by \eqref{eq:f_qmn}. (d--f) Coupling coefficients $\chi^{(m+n)}_{mn}$ for the dispersive nonlinear interaction, given by \eqref{eq:chi_mn}. For all three dispersion parameters shown here, $\beta_{2\text{s}}$ limits the phase-matching bandwidth, and we set $\beta_{2\text{s}} = \num{e-8}$, which corresponds to $\sim\num{e4}$ phase-matched comb lines for tens-of-femtosecond pulses with gigahertz repetition rates.
    }
    \label{fig:coupling-Fmn}
\end{figure*}

It is clear that the coupling functions $f_{mn}(\omega)$ play a crucial role in the physics of our model, as they affect the nonlinear Lindblad operators, the squeezing Hamiltonian under active pumping, and the dispersive nonlinear Hamiltonian. The physical considerations that govern the structure of $f_{mn}(\omega)$ are determined through the three-wave interaction integral \eqref{eq:coupling-fmn}, which in turn is dictated by the cavity mode profiles and the phase matching of the nonlinear interactions. To get intuition for the typical structure of $f_{mn}(\omega)$, we can consider some physical assumptions for both the cavity modes and the phase-matching conditions. We assume the signal modes, pump frequencies, and cavity dispersion are set consistently with the SPOPO model introduced in Sec.~\ref{sec:spopo}.

Let us also consider the special case of a one-dimensional cavity containing a section of nonlinear $\chi^{(2)}$ material with length $L$, and assume that the pump and signal modes propagate along the cavity optical axis $z$ with a one-dimensional phase front (i.e., they are mostly collimated) in the region where there is material nonlinearity. Then the mode functions can be written as
\begin{subequations}
\begin{align}
    \vect {D\sig}_m(\vect r) &= \vect {D\sig}_m(\vect r_\perp)e^{\im k_z({\omega\sig}_m) z} \\
    \vect {D\pump}_{\omega}(\vect r) &= \vect {D\pump}_{\omega}(\vect r_\perp)e^{\im k_z(\omega) z},
\end{align}
\end{subequations}
where $k_z(\omega)$ is the $z$ component of the wave vector and $\vect r_\perp$ is transverse to $z$. We also take the material nonlinearity to have a one-dimensional modulation due to quasi-phase-matching with period $k_\text{qpm}$~\footnote{To obtain the case of type-I phase matching instead, take $\sqrt{g_{mn}} \mapsto 2\sqrt{g_{mn}}, k_{\text{qpm}} \mapsto 0$.
}:
\begin{equation}
    \eta_{ijk}^{(2)}(\vect r) = \eta_{ijk}^{(2)}(\vect r_\perp) \cos(k_{\text{qpm}}z)
\end{equation}

Inserting these relations into \eqref{eq:coupling-fmn}, the function we are interested in can be written as~\footnote{Here, as throughout the paper, $\sinc x \coloneqq \sin x / x$.}
\begin{equation} \label{eq:Fmatrix-generic}
    f_{mn}(\omega) \approx g_{mn}^{1/2}(\omega)\sinc\Paren{\Phi_{mn}(\omega)},
\end{equation}
assuming $k_z(2\omega_0) - 2k_z(\omega_0) \gg 1/L$, with coupling rates $g_{mn}(\omega)$ and phase mismatch functions $\Phi_{mn}(\omega)$ given by
\begin{align}
    g_{mn}^{1/2}(\omega) &\coloneqq \frac{L}{2}  \int \dif^2 \vect{r}_\perp \, \eta^{(2)}_{ijk}(\vect r_\perp) \Paren{{{D^i\pump}\conj_{\omega}}{D^j\sig}_{\!m}{D^k\sig}_{\!n}}(\vect r_\perp) \label{eq:coupling-gmn} \\
    \Phi_{mn}(\omega) &\coloneqq \frac{k_{\text{qpm}} + k_z(\omega) - k_z({\omega\sig}_m) - k_z({\omega\sig}_n)}{2/L}. \label{eq:coupling-Phimn}
\end{align}

For the remainder of this paper, we make the further assumption that $g_{mn}(\omega)$ is approximately constant across the optical bandwidth of interest and we denote its value by $g_0$. Recalling \eqref{eq:pumped-spopo}, the nonlinear Lindblad operators and pump Hamiltonian for the SPOPO are then fully determined by the coefficients
\begin{equation}
    f_{mn} \coloneqq f^{(m+n)}_{mn} = g_0^{1/2}\sinc{\left(\Phi_{mn}\right)}, \label{eq:f_qmn}
\end{equation}
since $f^{(m+n)}_{mn} = f_{mn}({\omega\pump}_{m+n})$, and we similarly define
\begin{equation}
    \Phi_{mn} \coloneqq \Phi_{mn}({\omega \pump}_{m+n}).
\end{equation}

Finally, we also posit a form for the phase mismatch. By the quasi-degenerate nature of the SPOPO, we have $ 2k_z(\omega_0) - k_z(2\omega_0) = k_{\text{qpm}}$, or that $\Phi_{00} = 0$. Then, if the dispersion is sufficiently smooth within the optical bandwidth of interest, we can Taylor expand the wave-vector dispersion $k_z(\omega)$ to second order around the carrier frequencies of the fundamental ($\omega_0$) and second harmonic ($2\omega_0$). Thus, the phase-mismatch coefficients can be given the form
\begin{equation} \label{eq:coupling-Phimn-expanded}
  \Phi_{mn} \approx \beta_1(m+n) + \beta_{2\text{p}}(m+n)^2 - \beta_{2\text{s}}(m^2 + n^2),
\end{equation}
where $\beta_1 \coloneqq \frac 1 2 \Omega(\text{GVM})L$,  $\beta_{2\text{p}} \coloneqq \frac 1 4 \Omega^2 (\text{GDD}\pump)$, and $\beta_{2\text{s}} \coloneqq \frac 1 4 \Omega^2 (\text{GDD}\sig)$. Here the material dispersion parameters are GVM, the group velocity mismatch of pump relative to signal, and $\text{GDD}\pump$ ($\text{GDD}\sig$), the group delay dispersion of pump (signal), evaluated at $\omega_0$ for signal and $2\omega_0$ for pump. Under this simple model for the phase mismatch, we show in Figs.~\ref{fig:coupling-Fmn}(a--c) some typical forms for the dissipative nonlinear coefficients $f_{mn}$.

With this specific model for the phase matching, we can also obtain explicit expressions for the coupling rates $\chi^{(q)}_{mm'}$ in the nonlinear Hamiltonian $\hat H_\text{nl}$ as well. We only summarize the main arguments and results here, while the full derivation is provided in Appendix~\ref{app:nl_tensor}. As alluded to in Sec.~\ref{sec:io-theory-nl}, it is more convenient to calculate $\chi^{(q)}_{mm'}$ by directly calculating $\xi^{(q)}_{mm'}$ via \eqref{eq:xi-h} and \eqref{eq:memory-function}, rather than via the Cauchy principal-value integral (i.e., \eqref{eq:chi_cauchy_full}, or \eqref{eq:chi_cauchy_secular}); from \eqref{eq:xi_mm'}, we can take the imaginary part of $\xi^{(q)}$ to obtain $\chi^{(q)}$. As a result, the bulk of the calculation is in evaluating the memory functions, which by \eqref{eq:memory-function} have the form
\begin{align}
   &h_{mm'}^{(q)}(\tau) \coloneqq h_{m,q-m,m',q-m'}(\tau) = g_0\int_0^{\infty} \frac{\dif \omega}{2\pi} e^{\im({\omega \pump}_{q}-\omega)\tau} \nonumber\\ &\quad\qquad{}\times\sinc\Paren{\Phi_{m,q-m}(\omega)}
   \sinc\Paren{\Phi_{m',q-m'}(\omega)}.
  \label{eq:h_mm'}
\end{align}
Physically, the interaction among signal modes ${\omega\sig}_m$, ${\omega\sig}_{q-m}$, ${\omega\sig}_{m'}$, and ${\omega\sig}_{q-m'}$ is mediated by the pump reservoir, so this integral evaluates the total interaction by weighing the contributions at each reservoir frequency $\omega$. In this context, the exponential term produces oscillations for non-energy-conserving contributions, while the sinc functions account for momentum (i.e., phase) mismatch. In Appendix~\ref{app:nl_tensor}, we argue that for the evaluation of this integral, it suffices to expand the $\omega$ dependence of the phase mismatch to first order according to
\begin{align}
 \Phi_{m,q-m}(\omega) \approx \Phi_{m,q-m} + \frac 1 2 T_\text{nl}(\omega - {\omega \pump}_{q}),
\label{eq:Phi_series}
\end{align}
where $T_\text{nl}$ characterizes the time required for a pump photon of frequency ${\omega\pump}_q$ to travel through the nonlinear region. Inserting this expansion into \eqref{eq:h_mm'}, we can derive explicit expressions for $h^{(q)}_{mm'}$ (see Appendix~\ref{app:nl_tensor} for details), which turn out to vanish for memory times $\tau$ larger than order $T_\text{nl}$. Physically, this corresponds to the fact that an incident (virtual) pump photon from the reservoir can interact with the system signal modes for only the time $\sim T_\text{nl}$ during which it resides within the nonlinear region. As $T_\text{nl} \sim R_\text{fill}/\Omega$, where $R_\text{fill}$ is the ratio between the length of the nonlinear crystal and the total cavity length, the memory time is at most $\Omega\inv$, verifying that the memoryless Born-Markov master equation \eqref{eq:BM-master-eqn} is indeed self-consistent under the Markov condition for the SPOPO that $\Omega$ be sufficiently large.

Finally, inserting the explicit forms for $h^{(q)}_{mm'}$ into \eqref{eq:xi-h} and taking the imaginary part of $\xi^{(q)}_{mm'}$,
\begin{align}
    \chi^{(q)}_{mm'} &= \frac{g_0}{2} \Paren{\Phi_{m', q-m'}-\Phi_{m, q-m}}\inv \\
    &\qquad{}\times \bigl[\cos(\Phi_{m',q-m'})\sinc(\Phi_{m, q-m}) 
    \bigr. \nonumber\\
    &\qquad\qquad{}\bigl.- \cos(\Phi_{m,q-m})\sinc(\Phi_{m', q-m'})\bigr]. \nonumber
\end{align}
A particularly interesting two-dimensional slice of this three-dimensional tensor is $\chi^{(m+n)}_{mn}$, corresponding to Hamiltonian terms of the form $\chi^{(m+n)}_{mn} \hat s_{m}\dagg \hat s_{n}\dagg \hat s_m \hat s_{n}$ wherein the pairs of photons being created and destroyed are identical. In this case, we have the simplified expression
\begin{align}
    \chi^{(m+n)}_{mn} = \frac{g_0}{2\Phi_{mn}} \paren{\sinc(2\Phi_{mn})-1}.
    \label{eq:chi_mn}
\end{align}
Figures~\ref{fig:coupling-Fmn}(d--f) show these specific slices $\chi^{(m+n)}_{mn}$ for three different dispersion parameters. We see that the dispersive interaction tends to be weak where the dissipative interactions are strong and vice versa. This is physically intuitive as, due to its origin as an optical cascade, the dispersive interaction tends to be weak where the three-wave mixing is phase matched, which is also exactly where dissipative interactions due to two-photon loss (i.e., up-conversion to pump) are strong.

\section{Construction of supermodes}
\label{sec:supermodes}

Given the possibly large number of Lindblad operators and internal cavity modes, it is natural to ask whether there is a more efficient basis in which to describe the interactions. In general, one can apply any arbitrary basis transformation on the continuum pump modes to obtain new pump ``supermode'' operators
\begin{equation}
    \hat A^{(k)}_t \coloneqq \sum_q R_{kq} \hat a^{(q)}_t,
\end{equation}
where $R_{kq}$ is unitary, such that $[\hat A^{(k)}_t,\hat A^{\dagger(k)}_{t'}] = \delta_{kk'}\delta(t-t')$ in the white-noise limit. Neglecting any pumping terms for now, such a transformation can be applied to construct new supermode nonlinear Lindblad operators of the form
\begin{equation} \label{eq:nl-lindblads-R}
    \hat L'^{(k)}_\text{nl} \coloneqq \sum_q R_{kq} \hat L_\text{nl}^{(q)} = \sum_{m,n} \underbrace{R_{k,m+n} f_{mn}}_{f'^{(k)}_{mn}} \hat s_m \hat s_n.
\end{equation}
Whereas $\hat L^{(q)}_\text{nl}$ describes the coupling of the system to frequency modes of the reservoir at ${\omega\pump}_q$, $\hat L'^{(k)}_\text{nl}$ describes the coupling of the system to \emph{supermodes} of the reservoir.

At the same time, it is also possible to apply any arbitrary basis transformation on the signal modes, to define signal supermode operators
\begin{equation} \label{eq:Smodes}
    \hat S_{i} \coloneqq \sum_m T_{im} \hat s_m,
\end{equation}
where $T_{im}$ is also unitary, such that $[\hat{S}_i,\hat{S}^{\dagger}_j] = \delta_{ij}$. This can be used to generate a corresponding transformation of the signal reservoir operators into supermode reservoir operators
\begin{equation}
    \hat B^{(i)}_t \coloneqq \sum_m T_{im} \hat b^{(m)}_t,
\end{equation}
such that $[\hat B^{(i)}_t,\hat B^{\dagger(j)}_{t'}] = \delta_{ij}\delta(t-t')$ in the white-noise limit. Hence, $T_{im}$ can also be used to transform the linear Lindblad operators into supermode ones:
\begin{equation} \label{eq:loss-lindblads-T}
    \hat L'^{(i)}_\text{lin} \coloneqq \sum_m T_{im} \hat L^{(m)}_\text{lin} = \sum_m T_{im} \sqrt{2\kappa_m} \, \hat s_m.
\end{equation}

After performing these basis transformations, the supermode Lindblad operators can be furthermore written in terms of the signal supermodes as
\begin{subequations} \label{eq:lindblads-RT}
\begin{align}
    \hat L'^{(i)}_\text{lin} &= \sqrt{2} \sum_{j} \underbrace{\paren{\textstyle\sum_m \sqrt{\kappa_m}\, T_{im}T_{jm}\conj}}_{\sqrt{K}_{ij}} \hat S_j \label{eq:loss-lindblads-RT} \\
    \hat L'^{(k)}_\text{nl} &= \sum_{i,j} \underbrace{\paren{\textstyle\sum_{m,n} f'^{(k)}_{mn} T^*_{im}T^*_{jn}}}_{G^{(k)}_{ij}} \hat S_i \hat S_j. \label{eq:nl-lindblads-RT}
\end{align}
\end{subequations}
Similarly, the Hamiltonians corresponding to the detuning and dispersive nonlinearity can be written as
\begin{subequations}
\begin{align}
    \hat{H}_\mathrm{detuning} &= \sum_{i,j}\Delta_{ij}\hat{S}^{\dagger}_i\hat{S}_{j}, \quad\text{where} \\
    \Delta_{ij} &\coloneqq \sum_m \delta_m T_{im}T_{jm},
\end{align}
and
\begin{align}
    \hat{H}_\mathrm{nl} &= \sum_{i',j',i,j}J_{ i'j'ij}\hat{S}^{\dagger}_{i'}\hat{S}^{\dagger}_{j'}\hat{S}_{i}\hat{S}_{j}, \quad\text{where} \label{eq:Hnl_supermode}\\
    J_{i'j'ij} &\coloneqq \sum_q \sum_{m,m'} \chi^{(q)}_{mm'}T_{i'm'}T_{j',q-m'}T^*_{im}T^*_{j,q-m}.
\end{align}
\end{subequations}

The utility of this form is that, for physically realistic cases where $f_{mn}$ is relatively smooth (i.e., ``low-rank'') in the mode indices---as depicted in Fig.~\ref{fig:coupling-Fmn}, for example---the physics of the SPOPO is most concisely described using supermode bases for both pump and signal. In such cases, we can single out one nonlinear Lindblad operator to pump, while at the same time identifying a single signal supermode as the dominant degree of freedom in the cavity. The consequence is that in this supermode model, we need only consider a small range for the indices $(i,j,k)$ to accurately describe the physics, allowing us to truncate the interaction matrix $G^{(k)}_{ij}$, as well as the dispersive interaction tensor $J_{i'j'ij}$.

We remark that in performing these supermode transformations, it is numerically convenient to use the numerical technique described in Ref.~\cite[Sec.~5.2]{Patera2010} for downsampling the frequency comb indices (and performing appropriate rescalings of the SPOPO parameters) to take advantage of the smooth nature of $f_{mn}$ as a function of its indices. In this paper, we utilize such rescalings in calculating the supermode-basis coupling coefficients defined above.

\begin{figure}
    \includegraphics[width=1.0\linewidth]{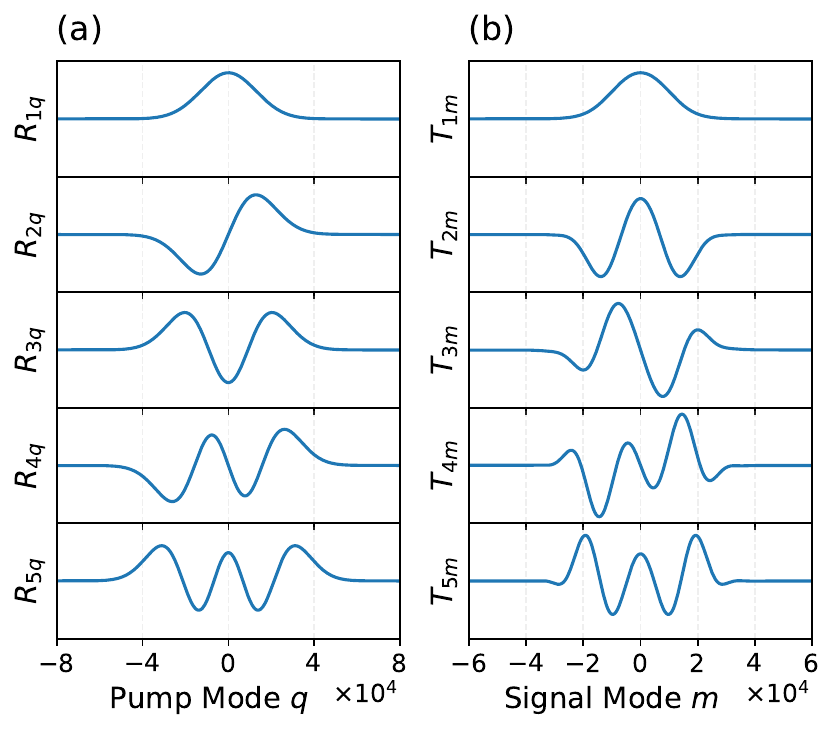}
    \caption{Supermode basis transformations on (a) pump reservoir modes and (b) signal system and reservoir modes, based on an SPOPO with dispersion and coupling parameters shown in Fig.~\ref{fig:coupling-Fmn}(c,f). In (a), the pump supermode basis $R_{kq}$ is chosen to be Hermite-Gaussian functions according to \eqref{eq:HG-modes} with $N\pump = \num{1.31e4}$, and in (b), the signal supermode basis $T_{im}$ diagonalizes $G^{(1)}_{ij}$ defined in \eqref{eq:nl-lindblads-RT}.}
    \label{fig:supermodes}
\end{figure}

To make the above discussion more concrete, consider the SPOPO with nonlinear coefficients $f_{mn}$ given by Fig.~\ref{fig:coupling-Fmn}(c). Suppose we would like to pump this SPOPO using a Gaussian spectrum centered on $2\omega_0$, with $N\pump$ comb lines spanning from the center to where the power falls to $1/e$. Then a good choice for pump supermodes are the Hermite-Gaussian functions:
\begin{equation} \label{eq:HG-modes}
    R_{kq} = \frac{1}{\Paren{\sqrt{\pi}N\pump 2^{k-1} (k-1)!}^{1/2}} H_{k-1}\Paren{q/N\pump} \e{-{\textstyle\frac12}(q/N\pump)^2},
\end{equation}
where $H_k$ is the physicists' Hermite polynomial of order $k$. This set of pump supermodes is illustrated in Fig.~\ref{fig:supermodes}(a).

Next we can choose the signal supermodes $T_{im}$ such that the matrix $G^{(1)}_{ij}$ is diagonal. As shown in Ref.~\cite{Patera2010} and below, this choice leads to a very simple form for the squeezing interaction induced by pumping in the $k=1$ supermode. This set of signal supermodes is shown in Fig.~\ref{fig:supermodes}(b). Note that $T_{im}$ is an orthogonal matrix since $f'^{(1)}_{mn}$ is real and symmetric for the interactions we are considering.

\begin{figure}
    \includegraphics[width=0.5\textwidth]{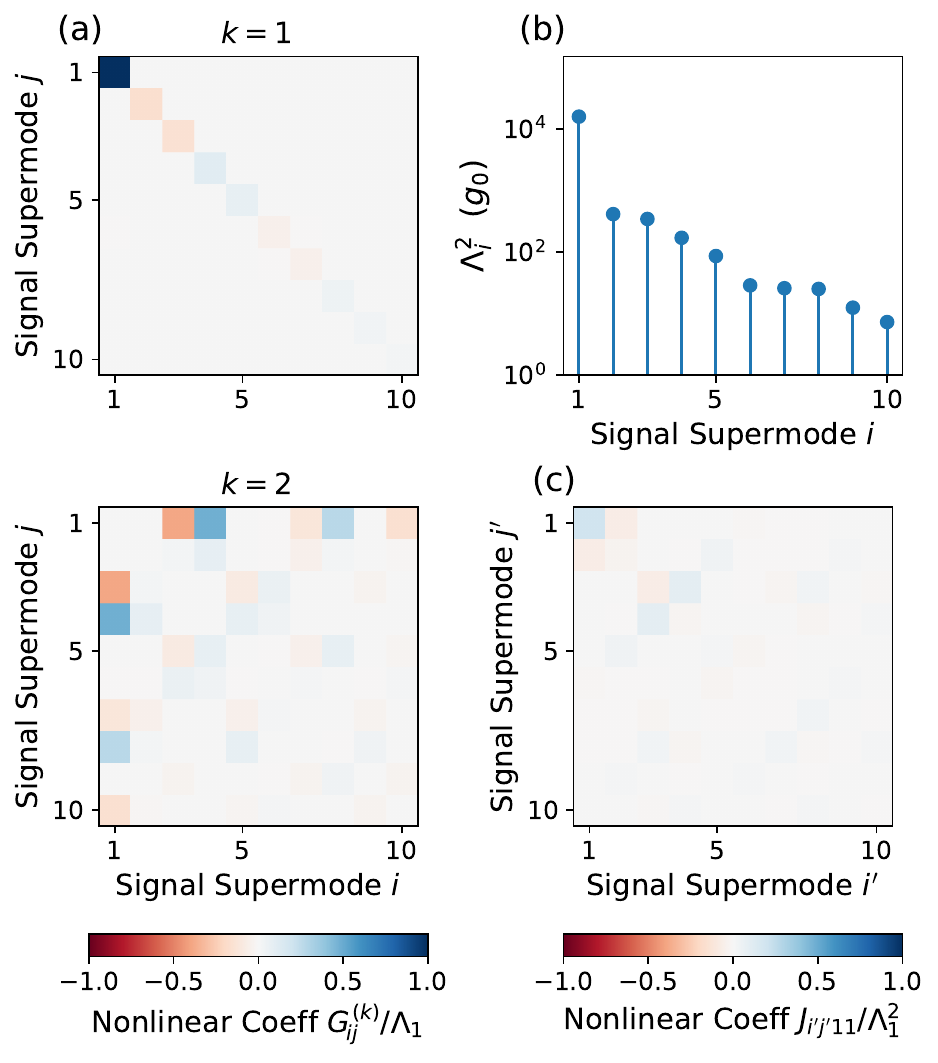}
    \caption{Structure of coupling coefficients in the supermode basis for an SPOPO with parameters shown in Fig.~\ref{fig:coupling-Fmn}(c,f), after performing the supermode-basis transformations shown in Fig.~\ref{fig:supermodes} and diagonalizing the matrix $G^{(1)}_{ij}$ as described in the text. The magnitudes $\Lambda_i^2$ of the first ten eigenvalues from the diagonalization are shown in (b), and $\Lambda_1$ is used to normalize all other coupling coefficients. (a) Coupling coefficients $G^{(k)}_{ij}$ (for $k = 1, 2$) that describe the nonlinear Lindblad operators $\hat L'^{(k)}_\text{nl}$. (c) Slice of the coupling coefficients $J_{i'j'11}$ that describes a dynamically relevant part (see discussion in main text) of the nonlinear dispersive Hamiltonian $\hat H_\text{nl}$.}
    \label{fig:nl-lindblads-RT}
\end{figure}

Pumping the SPOPO with amplitude $\mathcal A$ and a pump spectrum $\alpha^{(q)} = \mathcal A R_{1q}$ given by the first pump supermode, the supermode version of \eqref{eq:nl-Lindblads-displaced} gives
\begin{subequations} \label{eq:pumped-RT}
\begin{equation} \label{eq:nl-lindblads-pumped-RT}
    \hat L'^{(1)}_\text{nl} \mapsto \sum_{i,j} G^{(1)}_{ij} \hat S_i \hat S_j + \mathcal A = \sum_i \Lambda_i \hat S_i^2 + \mathcal A,
\end{equation}
while all the other Lindblad operators remain invariant. Furthermore, the squeezing Hamiltonian \eqref{eq:H-pumped-spopo} becomes
\begin{equation} \label{eq:Hsys-spopo-pumped-RT}
    \hat H_\text{pump} = \frac{\im\mathcal A}{2} \sum_i \Lambda_i \hat S_i^2 + \text{H.c.},
\end{equation}
\end{subequations}
where $\Lambda_i$ is the $i$th eigenvalue of $G^{(1)}_{ij}$; in this paper, eigenvalues are ordered in decreasing magnitude. This Hamiltonian describes a series of independent squeezing terms in the supermodes $\hat S_i$, in agreement with the physics obtained by Ref.~\cite{Patera2010} for an SPOPO below threshold after a supermode decomposition. 

In Fig.~\ref{fig:nl-lindblads-RT}, we carry out this supermode transformation and compute nonlinear coupling coefficients in the supermode basis for the SPOPO depicted in Fig.~\ref{fig:coupling-Fmn}(c,f) and using the supermodes shown in Fig.~\ref{fig:supermodes}. As expected from our construction of the signal supermodes $T_{im}$, the matrix $G^{(1)}_{ij}$ shown in Fig.~\ref{fig:nl-lindblads-RT}(a) is diagonal, with eigenvalues $\Lambda_i$ whose magnitudes are shown in Fig.~\ref{fig:nl-lindblads-RT}(b). The eigenvalues clearly show a rapid decay with increasing supermode index; consequently, by pumping the first supermode as described in \eqref{eq:pumped-RT}, the first supermode dominantly experiences most of the squeezing and gain in the system. When the SPOPO goes through threshold (i.e., experiences nonlinear dynamics), however, the action of the nonlinear Lindblad operators $\hat L^{(k)}_\text{nl}$, defined respectively by the coefficients $G^{(k)}_{ij}$, and shown for $k = 1,2$ in Fig.~\ref{fig:nl-lindblads-RT}(a), generally populate higher-order supermodes through \emph{multimode} nonlinear loss. Nevertheless, because the nonlinear interactions all tend to decay rapidly at larger supermode indices, excitations can remain confined to a relatively limited set of supermodes.

As shown in Fig.~\ref{fig:nl-lindblads-RT}(c), however, multi-supermode interactions can be mediated not only by the nonlinear loss but also by the coherent nonlinear dispersion due to $\hat H_\text{nl}$. One of the most important slices of the four-index tensor $J_{i'j'ij}$ is $J_{i'j'11}$, which acts on the state like $\hat S_{i'}\dagg \hat S_{j'}\dagg \hat S_1^2$, thus moving photons from the dominant supermode (index $i=1$) to higher-order ones (indices $i',j' > 1$). However, as is the case for the dissipative nonlinearity, Fig.~\ref{fig:nl-lindblads-RT}(c) shows that the coherent nonlinearity also decays at higher-order supermodes and furthermore are smaller in magnitude than the dominant dissipative nonlinear terms. This suggests it may be possible to tune parameters such as SPOPO dispersion or the pump spectrum in order to optimize the relative strengths of the dissipative and coherent nonlinear effects.

Finally, one of the most important features of ultrashort-pulse SPOPOs that can be seen from Fig.~\ref{fig:nl-lindblads-RT} is an effective \emph{pulsed enhancement} of the base nonlinear rate $g_0$, intuitively due to the temporal confinement of the field into a short pulse with higher peak power. Figure~\ref{fig:nl-lindblads-RT}(b) shows that all supermode coupling coefficients can be normalized to the principal eigenvalue $\Lambda_1$, which for this example takes on a value $\Lambda_1^2/g_0 \sim \num{e4}$. In the linearized regime where both the nonlinear Lindblad operators and the nonlinear Hamiltonian can be neglected, the dynamics cleanly decompose into independent squeezing on the supermodes $\hat S_i$ with squeezing rates enhanced by $\Lambda_i/\sqrt{g_0}$, physically corresponding to a reduction in the pump amplitude needed to reach threshold~\cite{Patera2010}. In the \emph{nonlinear regime}, multi-supermode interactions complicate the dynamics, but we still nevertheless expect the dynamical timescale for the mode $\hat S_1$, for example, to be on the order of $1/\Lambda_1^2$. In Sec.~\ref{sec:simulations}, we show dynamical quantum simulations of the supermode model presented in Fig.~\ref{fig:nl-lindblads-RT} that explicitly verify this fact, while in Sec.~\ref{sec:prospects}, we give some intuition for how this enhancement factor scales with experimental parameters and we discuss its implications for the realization of ultrashort-pulse SPOPOs in regimes of single-photon quantum nonlinearities.

\subsection{SPOPO equations of motion}
\label{sec:eoms}

After defining the supermodes for signal and pump, we can also transform the Heisenberg-Langevin equations of motion \eqref{eq:eoms_bare_freq} for the longitudinal modes into their corresponding supermode form as well. Assuming a coherent drive of amplitude $\mathcal A^{(k)}$ on each pump supermode $R_{kq}$,
\begin{align} \label{eq:eoms}
    \frac{\dif\hat S_{i'}}{\dif t} &= -\sum_i \paren{K_{i'i}+\im\Delta_{i'i}}\hat S_{i} - \sum_{k,j'} 2\mathcal A^{(k)} G^{(k)}_{i'j'} \hat S\dagg_{j'} \nonumber\\
    &{}- \sum_{j', i, j}
    \paren{\textstyle\sum_k G_{i'j'}^{(k)}G_{ij}^{(k)*} + 2\im J_{i'j'ij}}     \hat{S}^{\dagger}_{j'}\hat{S}_{i}\hat{S}_{j} \nonumber \\
    &{}- \sqrt2 \sum_{i} \sqrt{K}_{i'i} \hat B_{\text{in},t}^{(i)} - \sum_{k,j'}2G^{(k)}_{i'j'}{\hat S}_{j'}\dagg \hat A_{\text{in},t}^{(k)}, 
\end{align}
and the corresponding quantum input-output relations for the reservoir operators are:
\begin{subequations}
\begin{align}
    \hat B_{\text{out},t}^{(i)} &= \hat B_{\text{in},t}^{(i)} + \sqrt{2}\sum_j \sqrt{K}_{ij}\hat S_j \label{eq:lin-inout} \\
    \hat A_{\text{out},t}^{(k)} &= \hat  A_{\text{in},t}^{(k)} + \mathcal{A}^{(k)} + \sum_{i,j} G^{(k)}_{ij} \hat S_i \hat S_j, \label{eq:nl-inout}
\end{align}
\end{subequations}
where $\hat B_{\text{in},t}^{(i)}$ and $\hat  A_{\text{in},t}^{(k)}$ ($\hat{B}_{\text{out},t}^{(i)}$ and $\hat A_{\text{out},t}^{(k)}$) are input (output) quantum white-noise operators for the signal and pump, respectively. 

The first term in \eqref{eq:eoms} describes linear decay and coupling of the signal supermodes. In general, linear interactions can actually induce couplings between the different supermodes; this effect can potentially be useful for, e.g., designing couplings in coherent Ising machines~\cite{Wang2013}. A useful special case to consider is $\kappa_m = \kappa$, a constant for all $m$, and $\delta_m =0$ for all $m$ (evenly spaced signal modes); then this first term becomes $-\kappa \hat S_i$.

The second term in \eqref{eq:eoms} is the phase-sensitive OPO gain (phase-sensitive due to the dagger on $\hat S_j\dagg$), which is responsible for squeezing and thus the generation of signal excitation from pump driving. In general, this squeezing is multimode, but as discussed, we can choose to pump in the first supermode, so that $\mathcal A^{(1)} = \mathcal A$, a constant, with all others zero. Then by assuming that the signal supermode basis $T_{im}$ was chosen to diagonalize $\hat G^{(1)}_{ij}$ with eigenvalues $\Lambda_i$, this term becomes $-2\mathcal A \, \Lambda_i \hat S_i\dagg$.

The third term in \eqref{eq:eoms} describes the nonlinear interaction among different signal supermodes resulting from nonlinear interactions with the pump reservoir. Formally, this term (along with its associated noise term) is discarded in a linear treatment of SPOPOs~\cite{Patera2010}. Physically, this term provides a nonlinear clamping mechanism to stabilize the system when the gain produced by the second term exceeds the linear loss induced by the first term; this can be seen most readily in the single-supermode case where this term $\propto -(\hat S\dagg \hat S)\hat S$, signifying an intensity-dependent self-interaction. The real part of this term is a dissipative nonlinear interaction corresponding to two-photon loss, while the imaginary part corresponds to a coherent dispersive nonlinear interaction or, equivalently, an off-resonant cascaded $\chi^{(2)}$ interaction mimicking an effective four-wave-mixing $\chi^{(3)}$ interaction. For example, in the single-supermode case, the imaginary part $\propto -\im(\hat S^{\dagger}\hat S)\hat S$, corresponding to self-phase modulation. Notably, such a cascaded term is absent from a cw, \emph{single-frequency-mode} OPO under perfect phase matching, but it arises in the SPOPO case since phase-mismatched interactions are inherent to the multimode couplings described by $\chi_{mm'}^{(q)}$. In general, both the real and imaginary parts of this term are multimode and nonlinear: The supermode diagonalization procedure that cleanly simplifies the gain term still has residual structure which show up here and produces nonlinear couplings among supermodes.

The fourth and fifth terms in \eqref{eq:eoms} are quantum noise terms that describe the interaction between the signal supermodes and the supermodes of the two reservoirs: The first at the signal frequency band due to linear dissipation (which become independent interactions in the special case of $\kappa_m = \kappa$) and the second at the pump frequency band due to nonlinear parametric interactions. The reservoir modes themselves follow the input-output relations \eqref{eq:lin-inout} and \eqref{eq:nl-inout}, which also provide some insight into the physics of the SPOPO. In \eqref{eq:nl-inout}, we see that the nonlinear interaction with the cavity produces a two-photon contribution to the outgoing field, which can be thought of as broadband second-harmonic generation or, alternatively, a model for back-conversion of signal light back into pump light. Furthermore, when the sign of the back-conversion (i.e., the sign of $\sum_{ij} G_{ij}^{(k)} \hat S_i \hat S_j$) becomes out of phase with the drive amplitude $\mathcal A^{(k)}$, interference between the back-converted light and the input pump light manifests as pump depletion. In a linearized theory consistent with discarding the nonlinear terms of \eqref{eq:eoms}, the latter term in \eqref{eq:nl-inout} is omitted.

\section{The Markov condition for SPOPOs}
\label{sec:discussion-approx}

At a high level, our ability to formulate a quantum input-output theory for SPOPOs relies primarily on what we call the Markov condition, that both the bandwidth of the system-reservoir couplings $\kappa_m(\omega)$ and $f_{mn}(\omega)$ as well as the repetition rate $\Omega$ are both much larger than all other system dynamical rates. In conventional quantum input-output theory for cw (i.e., single-frequency-mode) cavities with high finesse, we routinely make the Markov approximation that the system-bath coupling is constant over a bandwidth larger than the rates of any system interactions. This facilitates derivation of a memoryless Born-Markov master equation for the system dynamics, input-output relationships for the reservoir dynamics, and even a quantum white-noise interpretation of the theory in terms of QSDEs. For example, when defining the reservoir operators
\begin{equation}
\hat b^{(\text{cw})}_t \coloneqq \int_{\mathcal B} \frac{\dif\omega}{2\pi} \, \hat b_\omega \e{-\im\omega t}, \quad \Sbrak{\hat b_\omega\dagg, \hat b_{\omega'}} = 2\pi\delta(\omega-\omega'),
\end{equation}
the band $\mathcal B$ is taken to have bandwidth greater than any system coupling rate (e.g., in the Hamiltonian or the linewidth). As a result, any excitations in these ``time-bin'' modes represented by $\hat b^{(\text{cw})}_t$ are very short compared to the time scale of the system dynamics, in which case we think of the interactions of these modes with the system to be independent, sequential events. Physically, the primary limitation to the bandwidth of $\mathcal B$ in most cw systems is the free spectral range (FSR) of the cavity: For an FSR of, say, \SI{100}{MHz}, numerical simulations produced by input-output theory are representative of the true physics down to the 10-ns scale, and it is rarely necessary to consider interactions occurring faster than this in cw systems.

As we have seen from Sec.~\ref{sec:input-output-theory}, the situation is more subtle for the pulsed OPO. To be concrete, let us consider the SPOPO with repetition rate $\Omega/2\pi$. As in the cw case, the bandwidth of $\mathcal S_m$ used to define $\hat b^{(m)}_t$ in \eqref{eq:sig-noise-operators} cannot exceed order $\Omega$ or else the partitions $\mathcal S_m$ will begin to overlap each other. As long as the system dynamical rates (e.g., in the few-photon excitation regime,  $|\mathcal{A}|^2$, $G_{i'j'}^{(k)*}G_{ij}^{(k)}$, $J_{i'j'ij}$, $K_{ij}$, $\Delta_{ij}$) are small compared to $\Omega$, we have shown that it is possible to formulate a quantum input-output theory encompassing both the linear and nonlinear couplings of the SPOPO to its environment. The end result is a \emph{multimode generalization} of the cw theory, in which, for example, the reservoir operators $\hat b^{(m)}_t$ and $\hat a^{(q)}_t$ are directly analogous to the operators $\hat b^{(\text{cw})}_t$, with the only complication being the frequency multiplexing of the reservoir spectrum into indexed partitions $\mathcal S_m$ and $\mathcal P_q$. In this sense, the Markov condition we have imposed is essentially a multimode version of the usual conditions needed to formulate quantum input-output theory in the single-mode case.

At the same time, for a pulsed system, it is arguably more natural to think about dynamics on a pulse-by-pulse basis, including dynamical effects that occur at sub-roundtrip timescales. However, such a picture cannot be faithfully captured by the input-output model we have presented. For example, we know from physical intuition that in an SPOPO without scattering losses, the pulse amplitude should only decrease when the pulse envelope hits the outcoupler; at single-roundtrip timescales, this is effectively a \emph{discrete} phenomenon. However, if we instantiate a coherent state of the first supermode $\hat S_1$ in the cavity and zoom into its dynamics on sub-roundtrip timescales, we see from the Heisenberg equations of motion \eqref{eq:eoms} that the expectation value of the field decays \emph{continuously} in time, contrary to physical intuition. As this effect is a natural consequence of formulating an input-output theory for pulsed OPOs, it is also a prominent feature of the model described in Refs.~\cite{DeValcarcel2006,Patera2010}.

The resolution to this discrepancy is the observation that if $\kappa \ll \Omega$ (which is imposed by the Markov condition), the dynamics of the system at \emph{longer timescales} of multiple roundtrips, e.g., the ringdown envelope of the pulses, are correctly reproduced. Thus, intuitively, our model after imposing the Markov condition holds only when the pulses do not experience dramatic changes over a single roundtrip or upon passing through a single optical element, a condition analogous to similar approximations made in classical pulsed nonlinear optics~\cite{Haus2000}. This approximation is a good characterization of pulsed OPOs with typical material nonlinearities and relatively low loss \footnote{It is worth noting that these conditions do not conflict with our numerical study of SPOPOs in the ``highly nonlinear'' regime, which merely involves the \emph{relative} scale of the nonlinear coupling rate to the linear dissipation rate.} such as those found in Refs.~\cite{Roslund2014,Ferrini2014}.

It is worth emphasizing that while this ``continuous-time approximation'' intuitively only requires that $\Omega$ dominate all other system dynamical rates, applying the approximation self-consistently is rather physically involved. As discussed in Sec.~\ref{sec:io-theory-nl}, the coupling of the cavity modes to the pump reservoir is not through a simple beamsplitter transfer function or scattering/loss spectrum, but rather through a \emph{phase-matching function} $f_{mn}(\omega)$ describing up-conversion to nonresonant pump photons, which generically can have richer spectral features and is more sensitive to experimental design. As a result, we cannot \textit{a priori} take the usual assumption that $f_{mn}(\omega)$ is flat as a function of $\omega$, i.e., that the coupling of modes $m$ and $n$ to the reservoir is memoryless. Rather, Appendix~\ref{app:nl_tensor} shows that the nonlinear interaction between signal and pump has an explicit memory time $\sim 1/\Omega$ (and possibly shorter if $R_\text{fill} < 1$), physically originating from the transit of a pump photon in the crystal. Thus the requirement for $\Omega$ to be sufficiently large is indirectly but intrinsically also responsible for ensuring that $f_{mn}(\omega)$ be sufficiently flat. In this sense, the continuous-time limit in which $\Omega$ is sufficiently large is the fundamental underlying assumption of our quantum SPOPO model.

In Appendix~\ref{app:propagation}, we explicitly relate the continuous-time dynamics of our model to a more conventional pulse-by-pulse description for the physics, by comparing their equations of motion in the classical, high-finesse limit where $\Omega$ is large. We therefore refer interested readers to Appendix~\ref{app:propagation} for details on how this continuous-time picture can be justified from within a pulse-propagation perspective.

Finally, as a topic for further research, we note that for high-gain, high-loss pulsed OPOs~\cite{Hamerly2016,Jankowski2018}, it is possible that quantum input-output theory would need significant modifications or even outright replacement in favor of a free-field formulation~\cite{Yanagimoto21, yanagimoto2021onset}, possibly involving the quantization~\cite{Drummond14,yanagimoto2020broadband, helt2020degenerate} of nonlinear classical field equations (such as the classical coupled-wave equations used in Appendix~\ref{app:propagation}), in order to model quantum pulse propagation at sub-roundtrip timescales.

\section{Numerical simulations}
\label{sec:simulations}

Having established the formalism of the model, we now turn to numerical simulations in order to explore the behavior of the nonlinear effects we have found in the model. In particular, we simulate SPOPOs in a highly nonlinear regime to study how multimode nonlinear quantum dynamics can affect important aspects of OPO phenomenology such as squeezing, output spectra, and non-Gaussian state generation. 

To perform numerical simulations, it is helpful to make various simplifying assumptions as mentioned throughout this paper. We consider an SPOPO as described in Sec.~\ref{sec:spopo}, with nonlinear interactions governed by the phase-matching assumptions made in Sec.~\ref{sec:phase-matching}; specifically, we consider the dispersion parameters shown in Fig.~\ref{fig:coupling-Fmn}(c). Following Sec.~\ref{sec:supermodes}, we recast the physics in supermode form, by pumping with strength $\mathcal A$ in the first Hermite-Gaussian pump supermode $R_{1q}$ in \eqref{eq:HG-modes}. We also choose the signal supermodes $T_{im}$ to diagonalize the interactions $G_{ij}^{(1)}$, with resulting eigenvalues $\Lambda_i$, corresponding to the eigenvectors $T_{im}$; these transformations correspond to those shown in Figs.~\ref{fig:supermodes} and \ref{fig:nl-lindblads-RT}. Finally, we assume that $\kappa_m \approx \kappa$ a constant, and all signal cavity modes have equal spacing $\Omega$ such that $\delta_m =0$ for all $m$, to simplify the linear dynamics as discussed in Sec.~\ref{sec:eoms}.

The result of these simplifying assumptions is a supermode quantum input-output model for an SPOPO with system Hamiltonian $\hat H_\text{pump} + \hat H_\text{nl}$ and a set of both linear and nonlinear Lindblad operators, given by
\begin{subequations} \label{eq:num-model}
\begin{align}
    \frac{\hat H_\text{pump}}{\kappa} &= \frac{\im r}{4} \sum_i \frac{\Lambda_i}{\Lambda_1} \hat S_i^2 + \text{H.c.} \label{eq:num-model-H} \\
    {\frac{\hat{H}_{\rm nl}}{\kappa}} &{= \eta \sum_{i,i',j,j'} \frac{J_{i'j'ij}}{\Lambda_1^2}\hat{S}_{i'}^{\dagger}\hat{S}_{j'}^{\dagger}\hat{S}_{i}\hat{S}_{j}} \label{eq:num-model-HJ} \\
    \frac{\hat L'^{(i)}_\text{lin}}{\sqrt\kappa} &= \sqrt{2} \hat S_i \label{eq:num-model-Lin}\\
    \frac{\hat L'^{(k)}_\text{nl}}{\sqrt\kappa} &= \sqrt\eta \sum_{i,j} \frac{G^{(k)}_{ij}}{\Lambda_1} \hat S_i \hat S_j + \frac{r}{2\sqrt\eta} \, \delta_{k1}, \label{eq:num-model-Lnl}
\end{align}
\end{subequations}
where $\Lambda_i = G_{ii}^{(1)}$ after diagonalization and we have introduced dimensionless parameters
\begin{equation}
    r \coloneqq \frac{2\mathcal A \Lambda_1}{\kappa} \quad\text{and}\quad
    \eta \coloneqq \frac{\Lambda_1^2}{\kappa}. \label{eq:parameters}
\end{equation}
The parameter $r$ is the pump parameter, representing the ratio of the pump field amplitude to the pump field amplitude at the (mean-field) threshold of the first supermode. The parameter $\eta$ is the ratio between the intensity-dependent and the linear decay rates of the first supermode in isolation. These dimensionless parameters incorporate the pump strength $\mathcal A$ and the scale $g_0$ of the nonlinear coupling coefficients. Thus in our simulations, we set $1/\kappa$ as the unit of time, and for a given $\eta$, we fix $g_0$ in Fig.~\ref{fig:nl-lindblads-RT}(b) such that $\Lambda_1 = \sqrt{\kappa\eta}$, while for a given $r$, we fix $\mathcal A = \kappa r / 2\Lambda_1$.

We perform numerical simulations in Julia using the QuantumOptics.jl package~\cite{Kramer2018}. We simulate quantum state evolution using standard numerical techniques, which we summarize in Appendix~\ref{app:methods} for convenience. We take advantage of the supermode decomposition to truncate the multimode simulation to a total of five signal modes (i.e., $1 \leq i \leq 5$) and to the first \num{20} nonlinear Lindblad operators (i.e., $1 \leq k \leq 20$). We use a Fock dimension of nine for the most dominant supermode and a dimension of three for the remaining higher-order supermodes. To check that our choices for numerical truncation are appropriate, we repeat calculations with increasing thresholds for truncation until we arrive at results that do not qualitatively change upon increase.

\begin{figure}
    \includegraphics[width=0.9\linewidth]{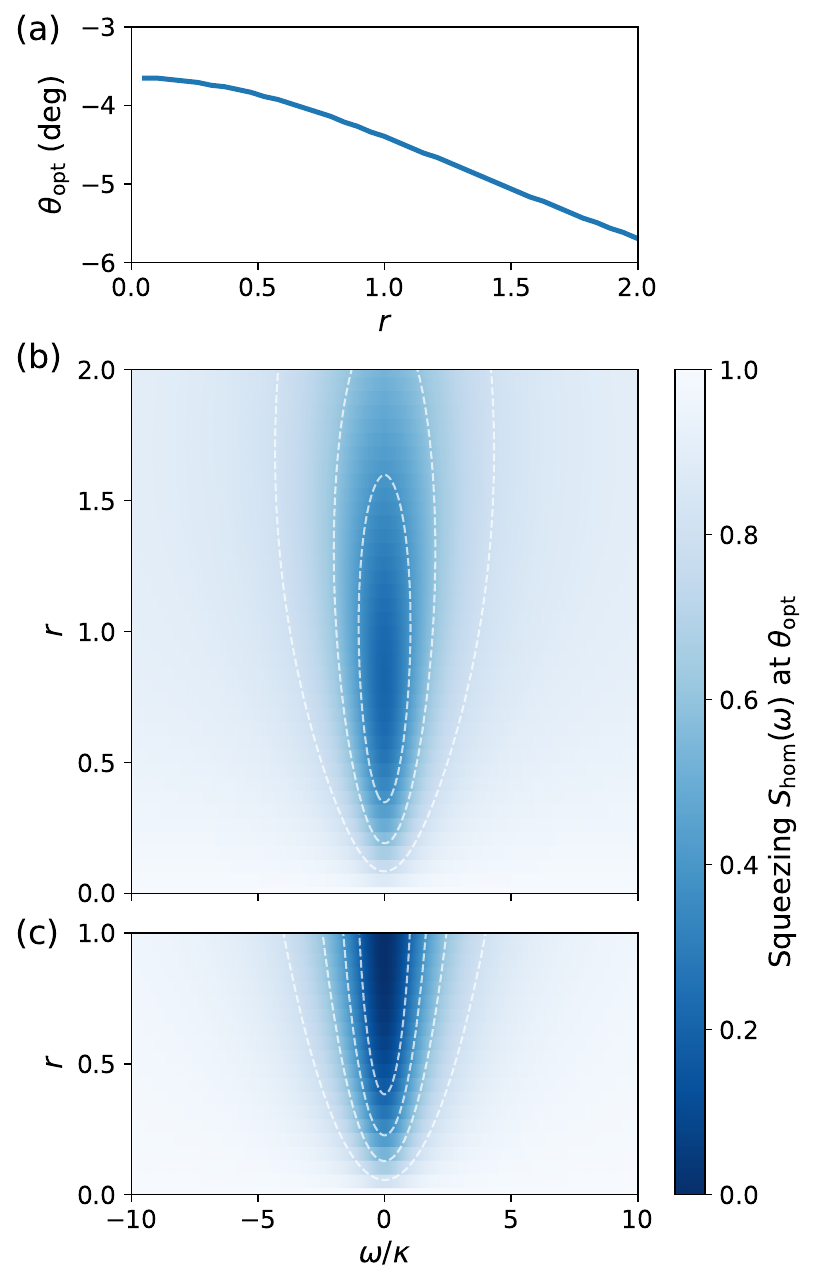}
    \caption{Squeezing generated by the first supermode $\hat S_1$ of an SPOPO \eqref{eq:num-model} operating in the quantum regime of $\eta = 1$, at varying pump parameter $r$. (a) Optimal squeezing angle $\theta_\text{opt}$ relative to the in-phase quadrature \eqref{eq:theta_opt}, induced by coherent nonlinear phase shifts in the model. (b) Steady-state squeezing spectrum $S_\text{hom}(\omega)$ \eqref{eq:S_hom}, measured at $\theta_\text{opt}$. (c) Analytic linearized approximation to the squeezing spectrum according to \eqref{eq:squeezing-lin}, neglecting both nonlinear loss and phase shifts. In (b) and (c), the vacuum level is normalized to 1. The dispersion parameters and choice of pump and signal supermodes follow Fig.~\ref{fig:nl-lindblads-RT}.}
    \label{fig:squeezing}
\end{figure}

\begin{figure}
    \includegraphics[width=0.95\linewidth]{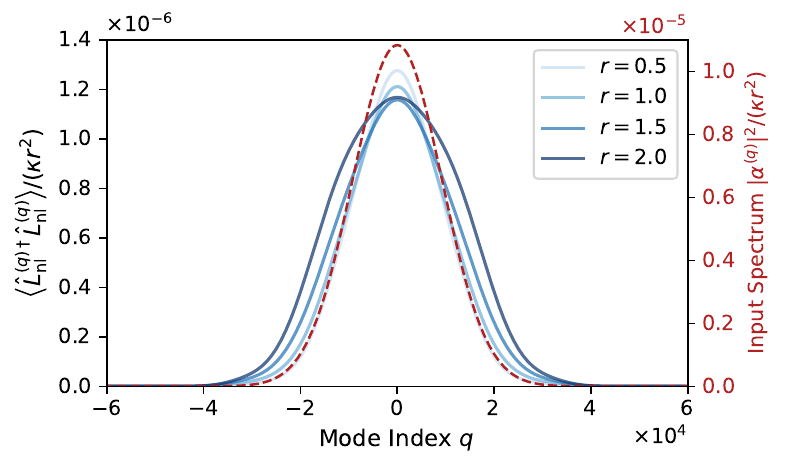}
    \caption{Steady-state input-output pump spectrum (i.e., photon-flux spectral density) of an SPOPO \eqref{eq:num-model} operating in the quantum regime of $\eta = 1$. The dashed line (right axis) corresponds to the input spectrum set by external pumping of the first pump supermode and is given by $|\alpha^{(q)}|^2/\kappa r^2 = R_{1q}^2/4\eta$; see also \eqref{eq:HG-modes}. The solid lines (left axis) show, at various pump parameters $r$, output spectra given by $\Vbrak{\hat L^{(q)\dagger}_\text{nl}\hat L^{(q)}_\text{nl}}/\kappa$. To better compare spectra at different pump parameters, we additionally normalize all curves by $r^2$.} 
    \label{fig:power}
\end{figure}

We first simulate the steady-state squeezing spectrum of the SPOPO in the first supermode in the highly nonlinear regime of $\eta = 1$. In this strongly quantum limit, our quantum model for the SPOPO exhibits important nonlinear deviations from conventional squeezing near and above threshold, arising from both the dissipative and coherent nonlinearities. In fact, the nonlinearity imparted by $\hat H_\text{nl}$ rotates the internal state of the SPOPO in phase space (see also Fig.~\ref{fig:wigner}), so we compute the squeezing not along a fixed quadrature, but instead along the angle of optimal squeezing, defined here as
\begin{equation}
    \theta_\text{opt} \coloneqq \arg\min_\theta \vbrak{\Paren{\hat{S}_1 e^{-\im \theta} + {\hat{S}}\dagg_1 e^{\im \theta}}^2},
    \label{eq:theta_opt}
\end{equation}
which produces the minimal variance in the resulting homodyne measurement. Figure~\ref{fig:squeezing}(a) shows how $\theta_\text{opt}$ changes with the pump parameter $r$, indicating that the dispersive nonlinearity can have a direct effect on the squeezing properties of an SPOPO.

At this optimal angle, Fig.~\ref{fig:squeezing}(b) shows, for the first signal supermode $\hat S_1$, the steady-state squeezing spectrum $S_\text{hom}(\omega)$ given by \eqref{eq:S_hom} (see Appendix~\ref{app:methods} for additional details). For comparison, we also show in Fig.~\ref{fig:squeezing}(c) the analytic squeezing spectrum
\begin{equation}
    S_\text{hom}^{(\text{lin})}(\omega) \coloneqq \frac{\omega^2 + \kappa^2(1-r)^2}{\omega^2 + \kappa^2(1+r)^2}, \label{eq:squeezing-lin}
\end{equation}
obtained by linearizing the equations of motion \eqref{eq:eoms} as done in Ref.~\cite{Patera2010} (corresponding to $\eta \rightarrow 0$). As we can see, the nonlinear model allows us to calculate squeezing beyond the threshold point of $r = 1$ where the linearized model breaks down. In addition, the spectra are markedly different in appearance: In this regime of high nonlinearity, both the bandwidth and the amount of squeezing are reduced, which is in accordance with the results of Ref.~\cite{Chaturvedi2002}. These results also suggest that, at $\eta \sim 1$, threshold as a mean-field concept is no longer sharply defined, as the mean photon number in the vacuum squeezed state ``below threshold'' ($r < 1$) is comparable to the mean photon number in the bright state ``above threshold'' ($r > 1$).

Another feature of the nonlinear quantum model is that the input-output behavior of the pump can be nontrivial. In Fig.~\ref{fig:power}, we show the steady-state optical output spectrum of the pump, also in the highly nonlinear regime of $\eta = 1$. Here, the spectrum is defined as $\Vbrak{\hat L^{(q)\dagger}_\text{nl}\hat L^{(q)}_\text{nl}}$ as a function $q$, i.e., the mean photon flux coming out of the SPOPO at frequency ${\omega\pump}_q$; this constitutes what one might measure, for example, on an optical spectrum analyzer. First, we find the output flux at nonzero $r$ (solid lines) is significantly lower than the input flux (dashed line; note the different vertical scales), indicating the presence of pump depletion. In fact, due to the strong nonlinearity associated with $\eta = 1$ and the ambiguity of classical threshold in this regime, pump depletion occurs even ``below threshold'' ($r < 1$), making undepleted-pump approximations generally invalid for quantum SPOPOs. Furthermore, as the pump parameter is increased, the output pump spectrum is nonlinearly distorted, since nonlinear interactions in the SPOPO increasingly excite higher-order signal supermodes with increasing $r$. These signal excitations back-convert into the pump reservoir through the action of the Lindblad operators $\hat L'^{(k)}_\text{nl}$.

\begin{figure*}[htp]
    \includegraphics[width=0.8\textwidth]{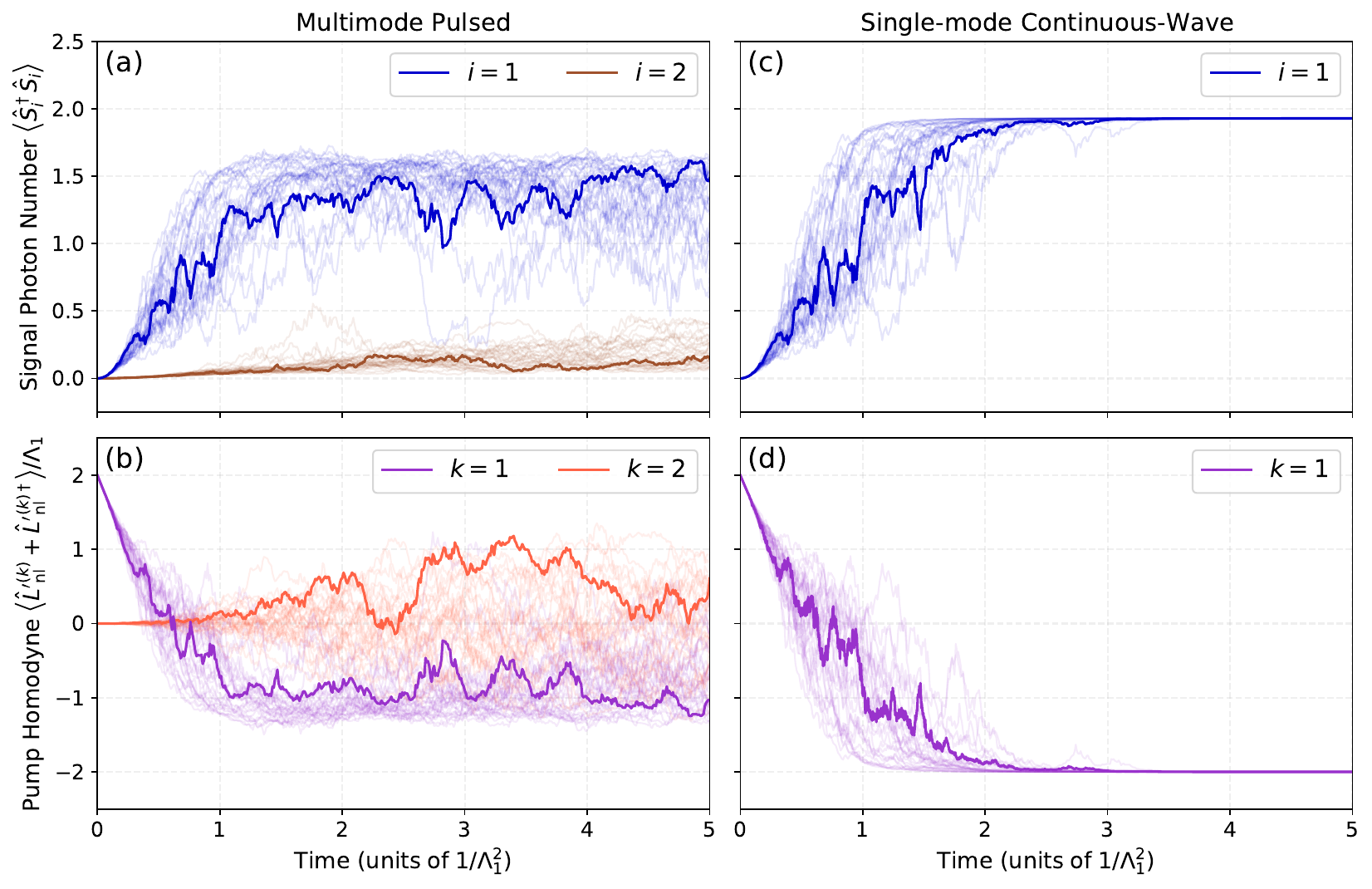}
    \caption{Quantum trajectory simulations comparing (a,b) the multimode SPOPO \eqref{eq:num-loseless-model} with (c,d) its single-mode cw counterpart, both evolving stochastically under no linear loss but continuous-time in-phase homodyne monitoring of the nonlinear loss channels $\hat L'^{(k)}_\text{nl}$. (a,c) Evolution of the internal photon number(s) in the signal supermode(s). (b,d) Evolution of the mean pump homodyne photocurrent. The dispersion parameters and choice of pump and signal supermodes follow Fig.~\ref{fig:nl-lindblads-RT}, and we use $p = 2$ for the pump parameter in \eqref{eq:num-loseless-model}. To facilitate direct comparison between the multimode and single-mode dynamics, bold trajectories indicate corresponding simulations that use the same seed in the random number generator.}
    \label{fig:stochastics}
\end{figure*}

To better understand the effects of the multimode nonlinearities in and of themselves, we also study the system in the absence of linear loss. This model can be obtained by a reparametrization of \eqref{eq:num-model}: We resubstitute $\eta$ and $r$ and cancel out $\kappa$ in favor of parametrizing time in $1/\Lambda_1^2$. More specifically, we use the model
\begin{subequations} \allowdisplaybreaks
\begin{align}
    \frac{\hat H_\text{{pump}}}{\Lambda_1^2} &= \frac{\im p}{4} \sum_i \frac{\Lambda_i}{\Lambda_1} \hat S_i^2 + \text{H.c.} \label{eq:num-lossless-model-H} \\
    {\frac{\hat{H}_\text{{nl}}}{\Lambda_1^2}} &{=  \sum_{i,j,i',j'} \frac{J_{i'j'ij}}{\Lambda^2_1} \hat{S}^{\dagger}_{i'}\hat{S}^{\dagger}_{j'}\hat{S}_{i}\hat{S}_{j}} \label{eq:num-lossless-model-HJ} \\
    \frac{\hat L'^{(k)}_\text{nl}}{\Lambda_1} &=  \sum_{i,j} \frac{G^{(k)}_{ij}}{\Lambda_1} \hat S_i \hat S_j + \frac p 2 \, \delta_{k1}, \label{eq:num-lossless-model-Lnl}
\end{align} \label{eq:num-loseless-model}
\end{subequations}
where we define a new pump parameter $p \coloneqq 2\mathcal{A}/\Lambda_1$.

We are especially interested in comparing this multimode model to its corresponding single-mode (or cw) version, which we define by restricting the indices appearing in \eqref{eq:num-loseless-model} to $i = j = k = 1$ (i.e., neglecting all higher-order pump and signal supermodes), as well as neglecting the self-phase modulation Kerr nonlinearity term $J_{1111}\hat{S}_1^{\dagger2}\hat{S}_1^2$, which vanishes for single-mode degenerate OPOs with perfect phase matching. The dynamics of such cw OPOs are well studied and relevant for applications of OPOs to quantum information processing; as shown in Ref.~\cite{Reid1993}, the steady state in the absence of linear loss is the pure cat state $\ket{\im \sqrt{p}} + \ket{-\im \sqrt{p}}$ (i.e., a superposition of coherent states).

\begin{figure*}[htp]
    \includegraphics[width=0.9\textwidth]{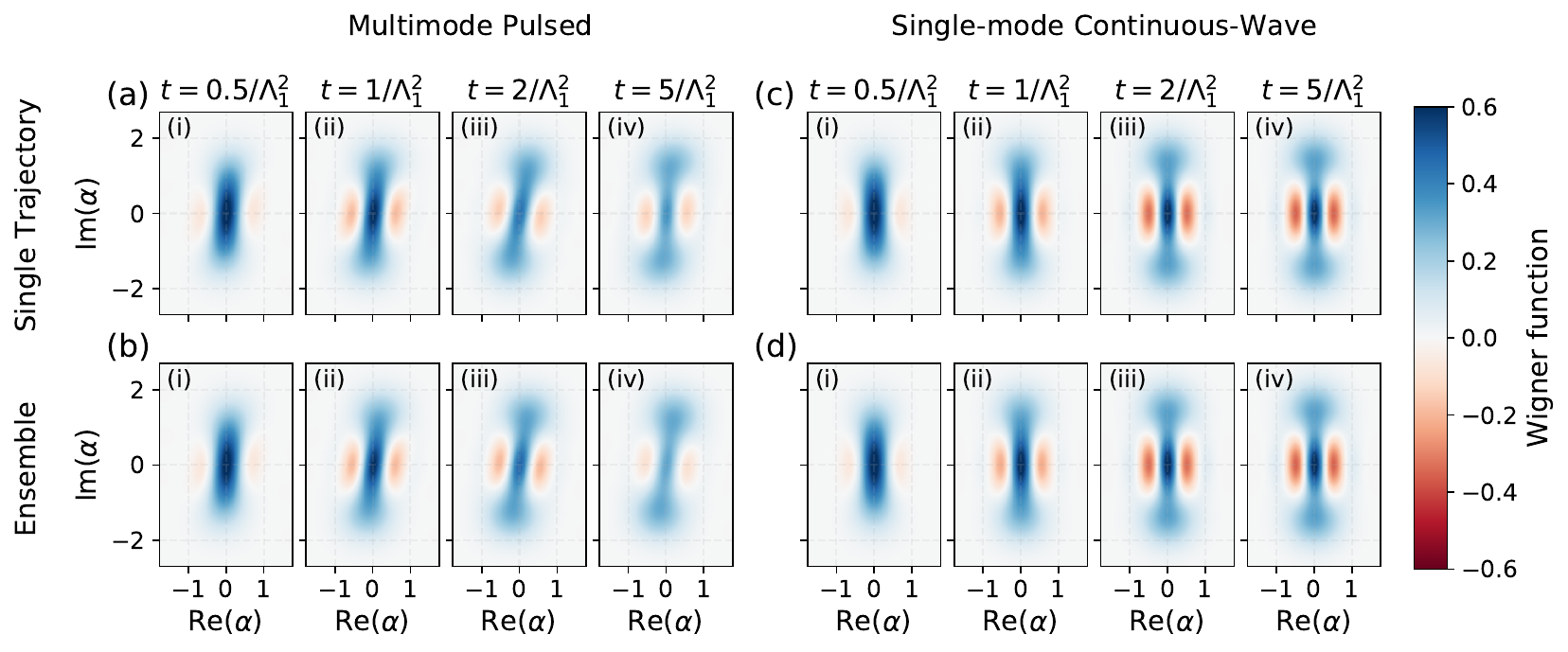}
    \caption{Wigner functions representing the state of the first signal supermode at various points in time (indexed with lowercase roman numerals); here (a,b) correspond to the multimode SPOPO, while (c,d) correspond to its single-mode counterpart. In (a) and (c), we show the Wigner functions for the bold stochastic trajectories in Fig.~\ref{fig:stochastics}(a) and (c), respectively. In (b) and (d), we show the Wigner functions of the ensemble average over all trajectories for each model, here computed via the unconditional master equation. Higher-order supermode components in the state of the SPOPO are partial-traced out.}
    \label{fig:wigner}
\end{figure*}

Figure~\ref{fig:stochastics} shows quantum trajectories from solving the stochastic Schr\"odinger equation for both an SPOPO and its corresponding single-mode cw OPO, at a pump parameter $p = 2$. As the only relevant Lindblad operators are the nonlinear ones, our choice of unraveling corresponds to in-phase homodyne monitoring of the \emph{pump} channels. The resulting trajectories provide direct evidence that multimode dynamics occur in the SPOPO, as the signal photon numbers in higher-order supermodes ($i > 1$) grow with time, while nonzero homodyne signals are observed in higher-order pump supermodes ($k > 1$). In the initial transient period ($t < 2/\Lambda_1^2$), the trajectories for the multimode SPOPO show stochastic variations qualitatively similar to those for the single-mode cw OPO. This is true in both the mean and variance of their respective ensembles, but also within a single instantiation of quantum noise, as is evident from the bold trajectories. After the transient period, however, the cw OPO quickly approaches steady state; specifically, it forms a pure single-mode cat state, which is dark to the nonlinear loss. On the other hand, the SPOPO continues to evolve stochastically due to still-increasing excitation in higher-order supermodes. These stochastic fluctuations about the ensemble mean decrease the purity of the ensemble, which indicates that the SPOPO experiences pump-induced decoherence even when the single-mode OPO has reached a pure state.

Similar conclusions can be reached by examining the quantum states generated in these trajectories, which we show in Fig.~\ref{fig:wigner} by plotting the Wigner functions of the first supermode at various points in time, both along the bold trajectories of Fig.~\ref{fig:stochastics} and for the ensemble-averaged dynamics (calculated from the unconditional master equation). As in Fig.~\ref{fig:stochastics}, we observe that the Wigner functions of the pulsed and cw OPO are qualitatively similar during the initial transient period (i.e., $t = 0.5/\Lambda_1^1, 1.0/\Lambda_1^2$). After this initial transient, the cw OPO quickly converges to its steady-state Wigner function, and the single trajectory looks similar to the ensemble average. On the other hand, for the SPOPO, the Wigner functions of the single trajectory differ from that of the ensemble average (e.g., they exhibit more negativity), indicating the presence of stochastic variations at later times. We furthermore observe that at later times in the \emph{single trajectory} of the SPOPO---for example, in Fig.~\ref{fig:wigner}(a)(iv)---the reduced state of the first supermode is \emph{also impure} (we estimate from simulations a purity of 0.6), even though the system state is pure under conditional evolution. This indicates that the multimode interactions in the SPOPO also entangle the various signal supermodes when conditioning on pump homodyne. Finally, it is also worth noting that the Wigner functions of the pulsed OPO are rotated compared to its cw counterpart: This is due to self-phase modulation of the first supermode, which is a consequence of the dispersive nonlinearity $\hat H_\text{nl}$, unique to the pulsed case.

\section{Prospects for experiments}
\label{sec:prospects}

In this section, we briefly discuss how our work relates to the experimental design of ultrashort-pulse SPOPOs, focusing in particular on prospects for observing the single-photon-regime nonlinear quantum behavior explored in Sec.~\ref{sec:simulations}. As discussed in Sec.~\ref{sec:discussion-approx}, the approximations we have made in our quantum model produce continuous-time dynamics, while in conventional models for experimental SPOPOs, outcoupling or losses are modeled using discrete beamsplitter operations and the evolution of the pulse through the nonlinear crystal is treated by directly integrating a set of coupled-wave equations. In Appendix~\ref{app:propagation}, we connect our model to the latter framework and derive an explicit mapping (in the mean-field limit) between the two models via a high-finesse, continuous-time approximation. 

Throughout this work, we have used a phenomenological frequency-independent coupling rate $g_0$ to characterize the $\chi^{(2)}$ nonlinearity (although a more \textit{ab initio} approach can start from \eqref{eq:Vnl} instead). This corresponds to assuming an instantaneous nonlinear coupling parameter $\epsilon$ in the coupled-wave equations of Appendix~\ref{app:propagation}; under such conditions, a figure of merit used in many experiments is the ``normalized second-harmonic conversion slope efficiency'', here defined as $\eta_0 \coloneqq (2\hbar\omega_0)\inv\epsilon^2$. This results in the correspondence
\begin{equation}
    g_0 = \frac{\hbar\omega_0}{2} \paren{\frac{L}{T}}^2 \eta_0,
\end{equation}
where $T$ is the roundtrip time of the cavity and $L$ is the propagation length through the nonlinear medium within the cavity. From a design perspective for nanophotonic cavities, it is often more natural to scale $L$ and $T$ together. In this case, we can re-express $L/T = R_\text{fill}v$, where $R_\text{fill}$ is the ratio between $L$ and the length of the cavity, and $v$ is the (average) group velocity of the signal pulse. Under this condition, the only dependence of $g_0$ on absolute lengths or the repetition rate arises via $R_\text{fill}$, which would ideally be a weak function thereof.

The decay rate $\kappa$ in our quantum theory can also be similarly related to experimentally relevant parameters. Following the derivation in Appendix~\ref{app:propagation}, we have
\begin{equation}
    \kappa = \frac{\ell^2}{2T},
\end{equation}
where $\ell^2$ denotes the fraction of power lost over one roundtrip due to linear mechanisms such as intrinsic losses and outcoupling. When this attenuation is dominated by scattering in a waveguide (rather than the outcoupler), $\ell^2$ also naturally scales with $T$, so we can express $\ell^2/T = \alpha_\text{loss} v$, where $\alpha_\text{loss}$ is the propagation power loss constant in units of \si{Np/[length]} (where Np stands for the neper). In this case, $\kappa$ does not scale with the absolute cavity length.

In Appendix~\ref{app:propagation} we also establish similar correspondences for the phase-mismatch coefficients $\Phi_{mn}$ and the pump amplitudes $\alpha^{(q)}$. We note again that these correspondences are only meaningful when the resulting dynamical rates in the supermode basis are sufficiently small compared to the repetition rate $\Omega$ (see Sec.~\ref{sec:discussion-approx}). That is, the resulting dynamics of the experimental SPOPO in question should have high finesse, and it should be valid to approximate its dynamics in continuous time over many roundtrips.

From an experimental and technological standpoint, one of the most important results from quantum studies of ultrafast SPOPOs such as this paper and Ref.~\cite{Patera2010} is the rigorous identification of a \emph{pulsed enhancement} of the nonlinear rate $g_0$, which intuitively arises due to the temporal confinement of the field to a short pulse. As first pointed out in Ref.~\cite{Patera2010} and discussed in Sec.~\ref{sec:supermodes}, decomposing the multimode quantum model into a supermode basis allows us to express the quantum model in terms of a small set of excited spectral-temporal supermodes. However, as we have shown in Sec.~\ref{sec:simulations}, nonlinear dynamics due to the two-photon loss and the dispersive optical cascade create interactions among the supermodes. This means that if we want to experimentally observe single-supermode behavior in the SPOPO while enjoying the pulsed enhancement due to temporal confinement, we must be careful to operate the SPOPO in a \emph{quasi-single-mode} regime. Identifying such a regime requires careful choice of the parameters of the system (especially for the dispersion), combined with a rigorous, concise model for its dynamics.

As discussed in Sec.~\ref{sec:supermodes} and borne out by quantum-dynamical simulations in Sec.~\ref{sec:simulations}, the first supermode $\hat S_1$ primarily experiences dynamics (e.g., squeezing rate and two-photon loss) on timescales set by $\Lambda_1$, where $\Lambda_i$ is the $i$th eigenvalue of the matrix $G^{(1)}_{ij}$ and is determined not only by the value of the underlying $g_0$ but also by the dispersion parameters of the system. Thus, a reasonable choice for quantifying the ``pulsed enhancement factor'' is the value of $\Lambda_1^2 / g_0$, \emph{assuming} we can operate the SPOPO in a quasi-single-mode regime. 

A detailed study systematically identifying such quasi-single-mode operating regimes, especially with all quantum effects taken into account, is beyond the scope of this paper and requires further research. However, for the purposes of establishing intuition for the experimental numbers relevant to the physics we explore in this paper, we can make some heuristic arguments as follows. For an SPOPO pumped in the first pump supermode as was done throughout Sec.~\ref{sec:simulations}, excitations are initially generated by the squeezing Hamiltonian $\hat H_\text{pump}$, which populates the signal supermodes via independent squeezing. As discussed in Sec.~\ref{sec:supermodes}, these excitations then facilitate the action of both the nonlinear dissipation and the nonlinear dispersion, which generically mixes the supermodes together in complicated ways. As a result, it is reasonable to heuristically consider the ratio $\Lambda_1^2 / \Paren{\textstyle\sum_i \Lambda_i^2} \sim 1$ as a necessary (but generically \emph{not sufficient}) condition for quasi-single-mode operation. In the following discussion, we use this ``single-modedness'' ratio as a heuristic metric for quasi-single-mode operation.

\begin{figure}
    \includegraphics[width=0.8\linewidth]{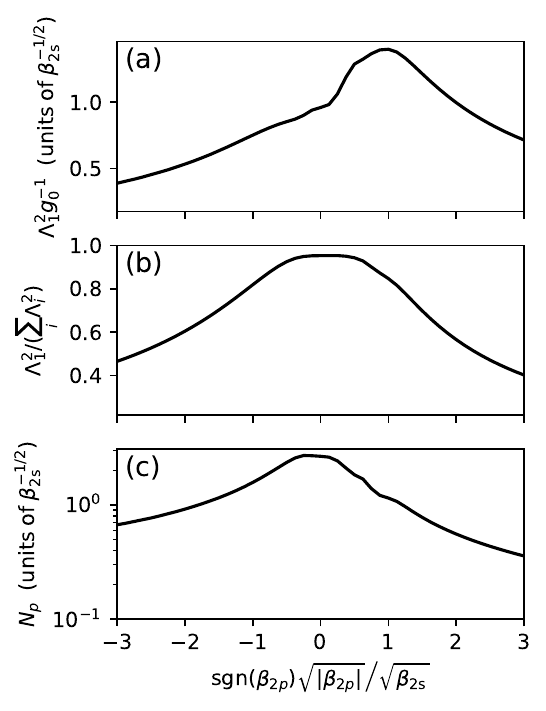}
    \caption{Heuristic SPOPO figures of merit under the second-order dispersion model \eqref{eq:coupling-Phimn-expanded}: (a) pulsed nonlinear enhancement factor (normalized by $\beta_{\text{2s}}^{-1/2}$) and (b) single-modedness ratio. These values are calculated using a number of pump comb lines $N\pump$ (see \eqref{eq:HG-modes}) chosen to maximize single-modedness, as shown in (c) (normalized by $\beta_{\text{2s}}^{-1/2}$).}
    \label{fig:dispersion}
\end{figure}

In Fig.~\ref{fig:dispersion}(a) and Fig.~\ref{fig:dispersion}(b), we show some representative plots for how the pulsed enhancement factor and the single-modedness ratio, respectively, change with some of the parameters used in our second-order dispersion model \eqref{eq:coupling-Phimn-expanded}. For simplicity, we focus here on the case where the GVM $\beta_1 = 0$; intuitively, we can expect quasi-single-mode operation to be difficult to achieve with large GVM. In this case, the GDD of the signal sets a characteristic scale for the number of cavity modes comprising the first supermode, allowing us to normalize by $\sqrt{\beta_{2\text{s}}}$ throughout. Furthermore, at each dispersion parameter, we have also chosen a particular pump bandwidth, defined by $N\pump$ in \eqref{eq:HG-modes}, in order to maximize the single-modedness; this chosen value of $N\pump$ is shown in Fig.~\ref{fig:dispersion}(c) for reference. For most parameters in the region where the single-modedness is high, $N\pump \sim 1/\sqrt{\beta_{\text{2s}}}$, indicating that the optimum width $\tau_\text{p}$ of the pump pulse is set approximately by the signal GDD, i.e., $\tau_\text{p}\Omega \sim 1/\sqrt{\beta_{2\text{s}}}$. Interestingly, we also see that slightly larger enhancements can be achieved by setting the pump GDD to be equal to the signal GDD, which is near the peak of Fig.~\ref{fig:dispersion}(a).

\begin{table*}[t]
    \centering
    \begin{tabular}{l l l l l | l}
        $\lambda_0$ (nm) & $\eta_0$ (\si{\percent\per\watt\per\centi\meter\squared}) & $g_0/2\pi$ (kHz) & $g_0/(\kappa = \SI{52}{MHz})$ & $g_0/\kappa \times \num{6710}$ & Notes \\
        \hline
        \num{2000} & \num{1000} & \num{0.018} & \num{2.2e-6} & \num{0.015} & Dispersion-engineered, Ref.~\cite{jankowski2020ultrabroadband} \\
        \num{1550} & \num{2600} & \num{0.060} & \num{7.2e-6} & \num{0.048} & CW only, ultra-low-loss, Refs.~\cite{wang2018ultrahigh,zhang2017monolithic} \\
        \num{913} & \num{33000} & \num{1.3} & \num{1.6e-4} & \num{1.0} & CW only, Ref.~\cite{park2021high}\\
        \num{775} & \num{120000} & \num{5.5} & \num{6.6e-4} & \num{4.5} & Theoretical, Ref.~\cite{jankowski2021dispersion}
    \end{tabular}
    \caption{Regimes of operation for high-finesse SPOPOs at various levels of nonlinearity, as measured by the SHG conversion efficiency $\eta_0$. For the loss rate, we use $\kappa = \SI{52}{MHz}$ based on a reference value of $\alpha_\text{loss} = \SI{3}{dB/m}$~\cite{zhang2017monolithic}; for more realistic values of $\alpha_\text{loss} \sim \SI{30}{dB/m}$, multiply $\kappa$ by \num{10}. The enhancement factor of \num{6710} is based on a reference waveguide resonator with length $L = \SI{1}{cm}$ and dispersion limited by signal GVD at \SI{1}{\fs\squared\per\mm} (see \eqref{eq:enhancement-reference}); more generally, this enhancement scales as $\sqrt{L/\text{GVD}\sig}$. We assume a cavity fill factor $R_\text{fill} = 1$ and average group velocity $v \approx c/2$.}
    \label{tab:numbers}
\end{table*}

Towards the goal of reaching the quantum limit of SPOPOs, we are especially interested in the results of Fig.~\ref{fig:dispersion}(a), which indicates that $\Lambda_1^2/g_0 \sim 1/\sqrt{\beta_{2\text{s}}}$, assuming $\beta_1$ is negligible. We also recall that $\beta_{2\text{s}} = \frac 1 4 \Omega^2 L (\text{GVD}\sig)$ where $\text{GVD}\sig \coloneqq k_z''(\omega_0)$. Combining these results and \emph{assuming quasi-single-mode operation}, we get that the pulsed enhancement can be estimated as
\begin{subequations}
\begin{align}
    \frac{\Lambda_1^2}{g_0} &\sim \frac{R_\text{fill}}{\pi} \sqrt{\frac{L}{v^2(\text{GVD}\sig)}} \\
    &\approx \num{6710} \times \sqrt{\frac{L}{\text{GVD}\sig} \times \frac{\SI{1}{\fs\squared\per\mm}}{\SI{1}{\cm}}} \label{eq:enhancement-reference},
\end{align}
\end{subequations}
where we have used $L/T = R_\text{fill} v$ in the first line, and assumed $R_\text{fill} \approx 1$ and $v \approx c/2$ in the second. In a nanophotonic platform, $g_0$ and (for a small outcoupler) $\kappa$ do not directly scale with $L$ as argued above, while $v$ and $\text{GVD}\sig$ are clearly independent of $L$. This expression provides an explicit expression for the pulsed enhancement we can expect in an ultrashort-pulse SPOPO, assuming it is operating in a quasi-single-mode regime in the limit of high finesse (and is GDD and scattering-loss dominated, etc.)  Finally, Table~\ref{tab:numbers} evaluates this expression for some representative numbers which have recently appeared in the literature on thin-film lithium niobate devices, to provide a sense of scale for the path forward to on-chip quantum SPOPOs.

\section{Conclusions}\label{sec:conclusions}
In this paper, we have investigated the nonlinear, multimode quantum effects in ultrashort-pulse OPOs, as revealed by a first-principles application of quantum input-output theory. We have rigorously derived a quantum input-output model for the SPOPO that features, under appropriate timescale conditions, a Born-Markov master equation in Lindblad form which includes the nonlinear interactions between the signal modes and the pump reservoir. The resulting model generically exhibits both dissipative nonlinearity, in the form of two-photon loss and modeled with nonlinear Lindblad operators, as well as a dispersive nonlinearity, in the form of an optical cascade and modeled with a four-wave-mixing Hamiltonian. By extending the supermode technique of Refs.~\cite{DeValcarcel2006, Patera2010}, we have obtained an efficient description for these multimode interactions. Numerical simulations using this supermode model show nonlinear phenomena such as pump depletion and back-conversion from signal to pump, as well as nonlinear corrections to the squeezing spectrum. In the regime of strong single-photon nonlinearities, we have found that, as expected, non-Gaussian physics in this regime can produce exotic intracavity states with Wigner-function negativity, but at the same time, the nonlinear multimode interactions, both dissipative and dispersive, play a significant and complicated role in the structure and dynamics of such states.

By inserting current as well as speculative experimental parameters into our model, we have seen that state-of-the-art devices in thin-film, dispersion-engineered nonlinear nanophotonics are now closer than ever before to reaching, in an all-optical platform, the quantum regime of few-photon operation. Such a breakthrough would make the on-chip short-pulse SPOPO a promising device for finally realizing the many quantum and coherent information processing applications proposed over decades of research into quantum OPOs, other quantum nonlinear oscillators, and networks thereof. At the same time, this paper also shows that despite this potential, a significant amount of work is still needed to model, control, and harness the multimode quantum effects that inherently arise from working with broadband devices like SPOPOs. For instance, the use of pump spectral shaping and/or advanced dispersion engineering may be needed to facilitate high-fidelity quantum operations and programmable quantum state generation. To enable such efforts in device and experimental design, further theoretical work will be needed to develop and apply more sophisticated model reduction techniques that can concisely and accurately capture the multimode quantum dynamics in the system.

\begin{acknowledgments}
The authors wish to thank Dodd J.\ Gray, Marc Jankowski, and Logan G.\ Wright, and Ryotatsu Yanagimoto for helpful discussions.

The authors wish to thank NTT Research for their financial and technical support. T.\,O., E.\,N., and H.\,M.\ acknowledge funding from NSF Award PHY-1648807 and from ARO Award W911NF-16-1-0086. R.H.\ was supported by an IC Postdoctoral Research Fellowship at MIT, administered by ORISE through U.S.\ DOE and ODNI. N.\,L.\ acknowledges funding from the Swiss SNSF and the NCCR Quantum Science and Technology. A.\,M.\ acknowledges funding from ARO Award W911NF-18-1-0285.
\end{acknowledgments}

\appendix

\section{Evaluating nonlinear interaction tensor} \label{app:nl_tensor}
In this section, we give details underlying the calculation of the coefficients $\chi^{(q)}_{mm'}$ that make up the nonlinear dispersive Hamiltonian, via direct evaluation of the memory function \eqref{eq:h_mm'} associated with the signal-pump interaction. We also provide some physical justification for the approximation \eqref{eq:Phi_series} used in performing these calculations.

Starting with \eqref{eq:h_mm'}, we first apply the linearization \eqref{eq:Phi_series} and then also assume we can extend the lower limit of the integral to $-\infty$, which is warranted as the integrand of \eqref{eq:h_mm'} is localized around $\omega = {\omega \pump}_q$. Under these assumptions, the memory functions can be expressed in terms of dimensionless quantities $\tilde \tau \coloneqq 2\tau/T_\text{nl}$ and $\tilde \delta \coloneqq \frac12 T_\text{nl}({\omega\pump}_q-\omega)$ as
\begin{equation}\label{eq:firstorderap}
    h_{mm'}^{(q)}(\tilde\tau) = \frac{g_0}{\pi T_\text{nl}} \int_{-\infty}^{\infty} \dif\tilde\delta\, Y_{qm}(\tilde\delta) Y_{qm'}(\tilde\delta) e^{\im\tilde\delta \tilde\tau},
\end{equation}
where $Y_{qm}(\tilde\delta)$ and its (inverse) Fourier transform $y_{qm}(\tilde\tau) \coloneqq \frac{1}{2\pi}\int_{-\infty}^{\infty} \dif\tilde\delta\, Y_{qm}(\tilde\delta)e^{\im\tilde\delta \tilde\tau}$ are given by
\begin{align}
    Y_{qm}(\tilde\delta) &= \sinc\paren{\Phi_{m}^{(q)} - \tilde\delta}, \label{eq:sinc-Y}\\
    y_{qm}(\tilde\tau) &= \frac{1}{2} \exp\paren{\im\Phi_{m}^{(q)}\tilde{\tau}} \Pi\paren{\tilde\tau},
\end{align}
where $\Pi(x)$ takes the value $1$ for $-1 < x < 1$ and is zero otherwise. For notational convenience, we have also introduced the notation $\Phi^{(q)}_m \coloneqq \Phi_{m,q-m}$. The convolution theorem then gives
\begin{align}\label{eq:h2}
    &h_{mm'}^{(q)}(\tilde\tau) = \frac{2g_0}{T_\text{nl}} \int_{-\infty}^{\infty} \dif\tilde\tau'\, y_{qm}(\tilde\tau') y_{qm'}(\tilde\tau-\tilde\tau') \\ 
    &\quad{}= \frac{g_0}{2T_\text{nl}}  e^{\im\Phi_{m'}^{(q)}\tilde{\tau}} \! \int_{-\infty}^{\infty} \dif\tilde\tau'\, e^{\im(\Phi_{m}^{(q)}-\Phi_{m'}^{(q)})\tilde{\tau}'}  \Pi\paren{\tilde\tau'} \Pi\paren{\tilde\tau-\tilde\tau'}. \nonumber
\end{align}
The last line shows that the memory function is given by a convolution between two temporal windows. For large $\tilde\tau$, the windows no longer overlap and the integrand vanishes for all $\tilde\tau'$. Consequently, 
\begin{align}
    h_{mm'}^{(q)}(\tau > T_\text{nl}) = h_{mm'}^{(q)}(\tilde\tau > 2) = 0.
\end{align}
For $\tau < T_\text{nl}$ where the two windows overlap, the integral evaluates to
\begin{align} \label{eq:h3}
  &h_{mm'}^{(q)}(\tau < T_\text{nl}) = \frac{g_0}{2T_\text{nl}} \frac{\im}{\Phi^{(q)}_{m'}-\Phi^{(q)}_m} \\
  &\qquad{}\times \paren{e^{\im(\Phi^{(q)}_m-\Phi^{(q)}_{m'})}e^{\im \Phi^{(q)}_{m'}\tilde \tau} -e^{\im(\Phi^{(q)}_{m'}-\Phi^{(q)}_m)}e^{\im \Phi^{(q)}_m\tilde \tau}}. \nonumber
\end{align}

Having derived an expression for the memory functions $h_{mm'}^{(q)}(\tau)$, we can now use \eqref{eq:xi-h} to find the nonlinear coupling coefficients by integrating over $\tilde\tau$. The result is
\begin{align}
    &\xi^{(q)}_{mm'} = \frac{g_0}{2}\sinc\Phi^{(q)}_m\sinc\Phi^{(q)}_{m'}
    + \im\frac{g_0}{2} \paren{\Phi^{(q)}_m-\Phi^{(q)}_{m'}}\inv \nonumber\\
    &\qquad{}\times \paren{\cos\Phi^{(q)}_{m'} \sinc\Phi^{(q)}_m - \cos\Phi^{(q)}_m \sinc\Phi^{(q)}_{m'}}.
    \label{eq:xi_qmm'_final}
\end{align}
We see that the real part of \eqref{eq:xi_qmm'_final} is consistent with the expression derived in the main text \eqref{eq:gamma_mm'} for the dissipative nonlinear couplings $\gamma^{(q)}_{mm'}$. But more importantly, the imaginary part of \eqref{eq:xi_qmm'_final} now provides us an expression for $\chi^{(q)}_{mm'}$, which specifies the coherent nonlinear couplings in $\hat H_\text{nl}$.

As an aside, since the nonlinear memory functions vanish for times $\tau \gg T_{\rm nl}$, and $T_{\rm nl} \sim R_{\rm fill}/\Omega$, we note that for small crystals relative to the cavity length where $R_{\rm fill} \ll 1$, it is possible to formulate a Markovian input-output master-equation model along the lines of Sec.~\ref{sec:io-theory-nl} which requires only the weaker condition that the rate of the system dynamics need only be less than $\Omega/R_{\rm fill}$, although this would require a larger binning of the input-output channels than we have chosen, in order to accommodate the larger interaction bandwidth. In this case, the requirement that the cavity remain high-finesse still stands, however.

\subsection{Validity of first-order expansion in $\delta$}
The function $\Phi_{m,q-m}(\omega)$ describing the phase mismatch between two signal modes at ${\omega\sig}_m$ and ${\omega\sig}_{q-m}$ with a pump photon at $\omega$ can be expanded more generally up to second order in the detuning $\delta = {\omega \pump}_q - \omega$ via
\begin{align}
    \Phi_{m,q-m}(\omega) \approx \Phi^{(q)}_m + \frac{1}{2}T_\text{nl} \delta + \frac{\beta_{2\mathrm{p}}}{\Omega^2} \delta^2,
    \label{eq:Phi_omega_expand}
\end{align}
where $T_\text{nl} \coloneqq k'_z({\omega\pump}_q) L$ is the time the pump photon spends in the nonlinear crystal, while $\beta_{2\text{p}}/\Omega^2 = k''_z({\omega\pump}_q)L$ under our second-order dispersion model \eqref{eq:coupling-Phimn-expanded}. The second-order term only becomes comparable to the first-order term when $|\delta/\Omega| \gtrsim \Omega T_\text{nl}/(2 \beta_{2\mathrm{p}}) \sim R_\text{fill}/\beta_{2\mathrm{p}}$, where $R_\text{fill}$ is the ratio of the nonlinear crystal length $L$ to the cavity length. For nanophotonic cavities, $R_\text{fill} \approx 1$, while for free-space SPOPOs, $R_\text{fill}$ is usually no smaller than \num{e-4}. On the other hand $1/\beta_{2\pump}^{1/2}$ represents the number of signal frequency modes within a characteristic frequency bandwidth set by the pump GDD; consequently, $\beta_{2\mathrm{p}}$ is very small for an ultrafast SPOPO (in this paper, we consider $\beta_{2\mathrm{p}}$ on the order of \num{e-8}). As a result, the second-order contribution is only significant for $|\delta|$ on the scale of many $\Omega$ (in this paper, $\sim\num{e8}\times\Omega$). Physically, we do not expect the pump reservoir at these large detunings away from ${\omega\pump}_q$ to mediate nonlinear interactions between signal frequencies at ${\omega\sig}_m$ and ${\omega\sig}_{q-m}$, and a first-order expansion in $\delta$ should suffice intuitively.

To develop the argument further, we see that in evaluating the memory functions \eqref{eq:h_mm'}, the \emph{most dominant} contributions to the integral occur when (i) the complex exponential is not too oscillatory (i.e., $\omega \approx {\omega\pump}_q$) and (ii) the phase matching is within the main lobe of the sinc functions (i.e., $|\Phi_{m,q-m}(\omega)| \lesssim \pi$). Physically, the nonlinear dispersive interaction between signal frequencies ${\omega\sig}_m$ and ${\omega\sig}_{q-m}$ are \emph{primarily mediated} by the pump reservoir at frequencies $\omega$ that are both (i) nearly resonant (i.e., energy conserving), and (ii) nearly phase matched (i.e., momentum conserving). Thus, in principle, a linear approximation to $\Phi_{m,q-m}(\omega)$ suffices if the \emph{effective integration bandwidth} of $\delta$ imposed by these two conditions is \emph{smaller} than the value of $\delta$ where the second-order contribution in \eqref{eq:Phi_omega_expand} becomes important.

As we can see in \eqref{eq:xi_qmm'_final}, analytic evaluation of the integral \eqref{eq:h_mm'} presupposing an expansion of $\Phi_{m,q-m}(\omega)$ to first order in $\delta$ suggests that the coupling constants are only nonzero when $\Phi^{(q)}_{m}$ and $\Phi^{(q)}_{m'}$ \emph{are also} nearly phase matched, i.e., within the main lobe of the sinc function. Combining this fact with the need for phase matching over $\delta$ in the integral (i.e., condition (ii) above), we can see (via \eqref{eq:sinc-Y}, for example) that we need $\delta \lesssim 2\Phi^{(q)}_{m}/T_\text{nl} \sim 2\Phi^{(q)}_{m'}/T_\text{nl} \sim T_\text{nl}\inv$. Thus, this argument suggests that the effective limits of the integral \eqref{eq:h_mm'} scale as $|\delta/\Omega| \lesssim 1/R_\text{fill}$. Finally, comparing this to the detuning at which the second-order term of \eqref{eq:Phi_omega_expand} becomes important, i.e., $|\delta/\Omega| \sim R_\text{fill}/\beta_{2\text{p}}$, we see that so long as $R_\text{fill} \gtrsim \sqrt{\beta_{2\text{p}}}$ (which is generically the case in most experiments, both on chip and in free space), then the first-order expansion in $\delta$ is consistent.

\section{Kramers-Kronig-like relationship between dissipative and dispersive part of nonlinear coupling constants}
\label{app:kk}

In this section we show how the real (giving rise to dissipation) and imaginary (giving rise to coherent cascade dynamics) parts of the coupling constants $\xi_{nmn'm'}$ are related by means of a Kramers-Kronig-like relationship, as a consequence of causality.

First, we start with the contribution to the nonlinear part of the master equation. Instead of the form~\eqref{eq:BM-master-eqn}, we begin with a master equation which only applies the Born approximation (separability of system and reservoir density matrices under the integral in~\eqref{eq:BM-master-eqn}), but not the Markov approximation that corresponds to the replacement $\hat{\rho}(t-\tau) \mapsto \hat{\rho}(t)$. This master equation, sometimes called the Nakajima-Zwanzig master equation, can be derived via projector operator techniques~\cite{Breuer2002}, and is given  by
\begin{equation}
    \mathcal L_\mathrm{nl} \hat{\rho} = -\int_0^{\infty} \!\dif\tau \tr_{\bar{\mathcal P}} \sbrak{\hat{V}_{\rm nl}(t), \Sbrak{\hat{V}_\mathrm{nl}(t-\tau),\hat{\rho}(t-\tau)\hat{\rho}_{\bar{\mathcal P}}}}.
\label{eq:B-master-eqn}
\end{equation}
Since both this equation and the time-convolutionless Born-Markov master equation in~\eqref{eq:BM-master-eqn} can be derived using second-order perturbation theory, the error associated with both equations can be expected to have the same scaling with the system-bath coupling~\cite{Breuer2002}. As such, while the Born-Markov form in~\eqref{eq:BM-master-eqn} is much more suitable for calculations, the Born form in~\eqref{eq:B-master-eqn} shows explicitly that the master equation theory has a causal structure, in that the evolution of the density matrix at time $t$ only depends on the state of the system at times $t' < t$.
Following similar steps to the derivation in Sec.~\ref{sec:input-output-theory}, we can write~\eqref{eq:B-master-eqn} in the form
\begin{align}
    \mathcal{L}_{\rm nl}\hat{\rho} = \sum_{m',n'}\sum_{m,n} &\int_{-\infty}^{\infty} \text{d}\tau h'_{mnm'n'}(\tau)\mathbb{L}_{mnm'n'}(t)\hat{\rho}(t-\tau) \nonumber \\ &+ \text{H.c.}
\end{align}
where $h'_{mnm'n'}(\tau) = h_{mnm'n'}(\tau)\theta(\tau)$, where $\theta(\tau)$ is the Heaviside step function, and we have defined the superoperator 
\begin{equation}
    \mathbb{L}_{mnm'n'}(t) \hat{\rho}  =  e^{\im({\omega\sig}_{m'} + {\omega \sig}_{n'}-{\omega\sig}_{m} - {\omega \sig}_{n})t}
    \Sbrak{\hat{s}_{m}\hat{s}_{n}\hat{\rho},\hat{s}^{\dagger}_{m'} \hat{s}^{\dagger}_{n'}}.
\end{equation}
In this form, it is clear that the evolution of the density matrix of the system depends on a set of bath memory functions $h'_{mnmn'}(\tau)$ which vanish explicitly for $\tau < 0$---a manifestation of causality, as the density operator at time $t$ can not depend on the state of the density matrix at time $t' > t$. 

We can then define a (frequency-shifted) Fourier transform of this function
\begin{align}\label{eq:xi_omega_def}
\xi_{mnm'n'}(\omega) &\coloneqq \int_{-\infty}^{\infty} \dif\tau h'_{mnm'n'}(\tau)e^{\im(\omega-{\omega \sig}_m - {\omega \sig}_n)\tau}.
\end{align}
If we assume $f_{mn}(\omega)$ to be a continuous and real function of $\omega$, we can apply the Sokhotski-Plemelj theorem for integrals over the real axis to find
\begin{subequations} \begin{align}\label{eq:xi_r&ia}
    \text{Re}\{\xi_{mnm'n'}(\omega) \} &= 
      \begin{cases}
      \frac{1}{2}f_{m'n'}(\omega)f_{mn}(\omega)  & \omega > 0 \\
                                   0 & \omega < 0 \\
  \end{cases}
    \\
    \text{Im} \{\xi_{mnm'n'}(\omega) \} &= -\text{P} \int_0^{\infty} \frac{\text{d}\omega'}{2\pi} \frac{f_{m'n'}(\omega')f_{mn}(\omega')}{\omega'-\omega}\label{eq:xi_r&ib}.
\end{align} \end{subequations}
Since $\xi_{mnm'n'}(\omega)$  is a (shifted) Fourier transform of a function $h'_{mnm'n'}(\tau)$ which vanishes for $\tau <0 $, by Titchmarsh's theorem~\cite{Toll1956Dec}, it is an analytic function of complex $\omega$ in the upper half plane. As such, we can use the residue theorem 
\begin{equation}\label{eq:contour1}
    \int_{\mathcal{C}} \text{d}\omega' \frac{\xi_{mnm'n'}(\omega')}{\omega'-\omega} = 0,
\end{equation}
for any contour $\mathcal{C}$ which is entirely located in the upper half plane. To derive the Kramers-Kronig-like relationship, we can choose a counter-clockwise contour which approaches from the interior of the contour a path consisting of a part along the real axis and a semicircle with radius $R \rightarrow \infty$; for well-behaved phase matching functions $f_{nm}(\omega)$, the contribution along this semicircle vanishes due to the exponential factor in~\eqref{eq:xi_omega_def}. We then evaluate~\eqref{eq:contour1} by using again the Sokhotski-Plemelj theorem, finding the Kramer-Kronig-like relations:
\begin{subequations} \begin{align}\label{eq:kk-a}
    \text{Re}\{\xi_{mnm'n'}(\omega) \} &= \frac{1}{\pi}\text{P}\int_{-\infty}^{\infty} \text{d}\omega' \frac{\text{Im} \{\xi_{mnm'n'}(\omega') \}}{\omega'-\omega}  \\
    \text{Im} \{\xi_{mnm'n'}(\omega) \} &= -\frac{1}{\pi}\text{P}\int_{-\infty}^{\infty} \text{d}\omega' \frac{\text{Re} \{\xi_{mnm'n'}(\omega') \}}{\omega'-\omega}.\label{eq:kk-b}
\end{align} \end{subequations}

We observe that the nonlinear coupling constants from Sec.~\ref{sec:input-output-theory} are related to these functions by $\xi_{mnm'n'}({\omega \sig}_m + {\omega \sig}_n) = \xi_{mnm'n'}$, and thus for the SPOPO model $\xi_{m,q-m,m',q-m'}({\omega \pump}_q) = \xi^{(q)}_{mm'}$. As such, we can conclude that the dissipative part of the nonlinear interaction---which gives rise to incoherent and nonunitary evolution of the system signal modes, and arises from real two-photon transitions---and the dispersive cascade part---which gives rise to coherent and unitary evolution of the system signal modes, and arises from virtual two-photon transitions---are not independent of each other, and indeed the existence of one implies the existence of the other. The result of this is that a theory of SPOPOs which includes nonlinear two-photon loss must also generally include the four-wave-mixing cascaded nonlinear Hamiltonian in order to preserve the causal structure of the interaction of the system signal modes with the pump reservoir.

\section{Correspondence with classical pulse-propagation theory of SPOPOs} \label{app:propagation}
In this section, we connect the mean-field limit of our quantum input-output theory for the SPOPO with a pulse-propagation model more commonly used in classical nonlinear optics. In the low-gain (i.e., short-crystal) limit, we show that the propagation of a classical pulse through a nonlinear $\chi^{(2)}$ crystal can be approximated as an input-output map; iterating this map over multiple roundtrips of the cavity in the low-loss limit yields continuous-time classical dynamics which correspond to the mean-field dynamics of the quantum theory. Notably, this derivation produces a mapping between classical figures of merit, such as the second-harmonic conversion efficiency, and the parameters used in the quantum model, thus conceptually bridging the theoretical results of this paper with more familiar models widely used in nanophotonic engineering. These results also provide an intuitive interpretation of the approximations made in the theory, connecting them to continuous-time approximations commonly employed in the classical domain to treat low-gain, high-finesse SPOPOs.

We begin by considering the mean-field dynamics generated by the quantum model given in Sec.~\ref{sec:spopo}. Starting from the quantum stochastic differential equations \eqref{eq:eoms_bare_freq} for the SPOPO, we take their expectation value and factor the resulting products using the mean-field approximation: 
\begin{subequations} \label{eq:mean-field-eoms}
\begin{align} 
    \frac{\dif\vbrak{\hat s_m}}{\dif t} = & -(\kappa_m + \im\delta_m) \vbrak{\hat s_m} - 2\sum_{q}f_{m, q-m}\alpha^{(q)}\vbrak{\hat s_{q-m}}^* \nonumber\\
    & -2\sum_q \sum_n \xi^{(q)}_{nm}  \vbrak{\hat s_{q-m}}^* \vbrak{\hat s_n} \vbrak{\hat s_{q-n}} 
\end{align}
Here a mean-field approximation was used to factor $\Vbrak{\hat s\dagg_{q-m}\hat s_n \hat s_{q-n}}$ into $\vbrak{\hat s_{q-m}}^* \vbrak{\hat s_n} \vbrak{\hat s_{q-n}}$. This approximation is justified if the internal state of the SPOPO is a multi-mode coherent state, e.g., in the classical regime where loss is much larger than nonlinearity.

In addition, the input-output relationships for the signal and pump channels reduce to
\begin{align}
    \Vbrak{\hat b^{(m)}_{\text{out}}} &= \sqrt{2\kappa_m} \vbrak{\hat s_m} \\
    \Vbrak{\hat a^{(q)}_{\text{out}}} &= \alpha^{(q)} + \sum_m f_{m,q-m} \vbrak{\hat s_m} \vbrak{\hat s_{q-m}} ,
\end{align}
\end{subequations}
since $\Vbrak{\hat b^{(m)}_{\text{in}}} = \Vbrak{\hat a^{(q)}_{\text{in}}}=0$. Thus, we see that the quantum theory consists of only the parameters: $\kappa_m$, describing the linear loss rates; $\alpha^{(q)}$, describing the pump flux amplitudes; and $f_{mn}$ and $\xi_{mn}^{(q)}$, representing the nonlinear interactions in the system.

We now introduce a simple but typical description of classical SPOPOs, which uses a set of coupled-wave equations to model pulse propagation of pump and signal through the crystal and an iterative loop to model the recycling of the signal within the cavity. In the mostly-collimated, one-dimensional cavity situation considered in Sec.~\ref{sec:phase-matching}, the classical electric fields of the signal and pump can be written as
\begin{subequations} \begin{align}
    \vect E\sig(t;\vect r) &= \vect E\sig(\vect r_\perp)\e{\im k(\omega_0)z}\e{-\im\omega_0 t} {\alpha\sig}(t;z) + \text{c.c.} \\
    \vect E\pump(t;\vect r) &= \vect E\pump(\vect r_\perp)\e{\im k(2\omega_0)z}\e{-2\im\omega_0 t} {\alpha\pump}(t;z) + \text{c.c.},
\end{align} \end{subequations}
where $\vect E\sig$ and $\vect E\pump$ are some suitable mode functions, $k(\omega)$ is the wavevector component as a function of frequency along the propagation direction $z$, and $\omega_0$ is the signal carrier frequency chosen such that ${\alpha\sig}(t)$ and ${\alpha\pump}(t)$ are envelope functions for the signal and pump fields satisfying the slowly-varying envelope approximation. The signal envelope can be related to the Fourier mode amplitudes of a cavity with roundtrip time $T = 2\pi/\Omega$ via
\begin{subequations} \label{eq:classical-mode-amplitudes}
\begin{equation}
    \alpha\sig(t;z) = \frac{1}{\sqrt T} \sum_m {\alpha\sig}_m(z)\e{-\im m\Omega t}.
\end{equation}
Although the spectrum of the pump envelope $\alpha\pump(t)$ can technically be continuous, if we assume that the frequencies $2\omega_0 + q\Omega$ defined by the time window $T$ are sufficiently fine to approximate that spectrum (or indeed, if the system is pumped with a mode-locked laser), then we can also similarly write
\begin{equation}
    \alpha\pump(t;z) = \frac{1}{\sqrt T} \sum_q {\alpha\pump}_q(z)\e{-\im q\Omega t}.
\end{equation}
\end{subequations}
The coupled-wave equations which describe the evolution of the envelope functions within the crystal are
\begin{subequations} \label{eq:coupled-wave-equations}
\label{eq:cwe}
\begin{align}
    \partial_z\alpha\sig(t;z) &= \im k\sig(\im\partial_t) \alpha\sig(t;z) + \epsilon (\alpha\pump \alpha\sig\conj)(t;z) \\
    \partial_z\alpha\pump(t;z) &= \im k\pump(\im\partial_t) \alpha\pump(t;z) -\frac 1 2 \epsilon \alpha\sig^2(t;z),
\end{align}
\end{subequations}
where we have introduced $k\pump(\bar\omega) = k(2\omega_0+\bar\omega) - k(2\omega_0)$ and $k\sig(\bar\omega) = k(\omega_0+\bar\omega) - k(\omega_0)$, allowing us to write the dispersion operators as formal power series---e.g., $k\sig(\im\partial_t) = \sum_d (1/d!) k\sig^{(d)}(0)(\im\partial_t)^d$, where $k\sig^{(d)}$ denotes the $d$th derivative of $k\sig$. PDEs of this form can be readily solved numerically, e.g., via Fourier split-step methods.

To bring the classical dynamics closer to the quantum formalism, we turn this pair of PDEs into a discrete set of ODEs by moving to the Fourier domain using \eqref{eq:classical-mode-amplitudes}. We also move into an ``interaction frame'' via
\begin{subequations} \label{eq:interaction-frame}
\begin{align}
    {\widetilde\alpha}{{}\sig}_m(z) &= \e{-\im k\sig(m\Omega)z} {\alpha\sig}_m(z) \\
    {\widetilde\alpha}{{}\pump}_q(z) &= \e{-\im k\pump(q\Omega)z} {\alpha\pump}_q(z).
\end{align}
\end{subequations}
With both of these transformations, the coupled-wave equations simplify to (suppressing $z$ arguments)
\begin{subequations} \label{eq:coupled-wave-equations-IF}
\begin{align}
    \frac{\dif{\widetilde\alpha}{{}\sig}_m}{\dif z} &= \frac{\epsilon}{\sqrt{T}} \sum_q {\widetilde\alpha}{{}\pump}_q{\widetilde\alpha}{{}\sig}_{q-m}\conj \e{+\im\Delta k_{m,q-m}z} \\
    \frac{\dif{\widetilde\alpha}{{}\pump}_q}{\dif z} &= -\frac12\frac{\epsilon}{\sqrt T} \sum_m {\widetilde\alpha}{{}\sig}_m{\widetilde\alpha}{{}\sig}_{q-m}\e{-\im\Delta k_{m,q-m}z},
\end{align}
\end{subequations}
where we have defined the momentum mismatch to be
\begin{equation}
    \Delta k_{mn} = k\pump\paren{m\Omega + n\Omega} - k\sig\paren{m\Omega} - k\sig\paren{n\Omega}.
\end{equation}
Our task is now to integrate these equations of motion through a crystal of length $L$. For simplicity, we specify the input facet of the crystal to be at $z = -L/2$, with the output facet at $z = L/2$.

For both signal and pump, we also henceforth denote $\widetilde\alpha = \widetilde\alpha(z=-L/2)$ and $\widetilde\alpha' = \widetilde\alpha(z=+L/2)$ to suppress the arguments at the facets, and we reserve $\widetilde\alpha(z)$ to denote the field only in the interior (i.e., when $-L/2 < z < L/2$).

In a limit where the single-pass gain is sufficiently weak, we can approximately solve these equations of motion via a Picard iteration. Let the Picard iterates for signal and pump be $\widetilde\alpha\sig^{(i)}(z)$ and $\widetilde\alpha\pump^{(i)}(z)$. To determine the number of iterations we should perform, we introduce a parameter $\delta$ such that
\begin{equation}
    \frac{\epsilon L}{\sqrt{T}} \sim \alpha\pump \sim \delta^{1/2},
\end{equation}
and we perform the Picard iteration until further iterations no longer give any corrections of order $\mathcal O(\delta)$---that is, we want to derive the dynamics via Picard iteration up to first order in $\delta$, assuming $\alpha\sig$ is zeroth order in $\delta$.

As usual, the zeroth iteration is simply given by $\widetilde\alpha\sig^{(0)} = \alpha\sig$ and $\widetilde\alpha\pump^{(0)}(z) = \alpha\pump$. The first iteration is then
\begin{subequations} \label{eq:picard-order-one}
\begin{align}
    {\widetilde\alpha}{{}\sig}_m^{(1)}(z) &= {\widetilde\alpha}{{}\sig}_m + \frac{\epsilon L}{\sqrt{T}} \sum_q {\widetilde\alpha}{{}\pump}_q {\widetilde\alpha}{{}\sig}\conj_{q-m} I_{mq}(z) \\
    {\widetilde\alpha}{{}\pump}_q^{(1)}(z) &= {\widetilde\alpha}{{}\pump}_q - \frac12\frac{\epsilon L}{\sqrt{T}} \sum_m {\widetilde\alpha}{{}\sig}_m {\widetilde\alpha}{{}\sig}_{q-m} I\conj_{mq}(z),
\end{align} \end{subequations}
where we have introduced
\begin{equation}
    I_{mq}(z) \coloneqq \int_{-L/2}^z \exp\paren{\im\Delta k_{m,q-m}z'} \frac{\dif z'}{L}
\end{equation}
Inspecting the second terms of \eqref{eq:picard-order-one}, we see that they are $\sim\delta$ and $\sim\delta^{1/2}$ for signal and pump, respectively. The next Picard iteration gives
\begin{align}
    &{\widetilde\alpha}{{}\sig}_m^{(2)}(z) = {\widetilde\alpha}{{}\sig}_m^{(1)}(z) - \frac{\epsilon^2 L^2}{4T} \sum_{q,n} {\widetilde\alpha}{{}\sig}\conj_{q-m}{\widetilde\alpha}{{}\sig}_{n}{\widetilde\alpha}{{}\sig}_{q-n} H_{mnq}(z) \nonumber\\
    &\qquad{} +\mathcal O\Paren{{\textstyle\frac{\epsilon^2 L^2}{T}\widetilde\alpha\pump\conj \sim \delta^{3/2}}} + \mathcal O\Paren{{\textstyle\frac{\epsilon^2 L^2}{T}|\widetilde\alpha\pump|^2 \sim \delta^2}} \nonumber \\
    &{\widetilde\alpha}{{}\pump}_q^{(2)}(z) = {\widetilde\alpha}{{}\pump}_q^{(1)}(z) + \mathcal O\Paren{{\textstyle\frac{\epsilon^2 L^2}{T}\widetilde\alpha\pump \sim \delta^{3/2}}} \\
    &\qquad{} + \mathcal O\Paren{{\textstyle\frac{\epsilon^3 L^3}{T^{3/2}}\widetilde\alpha\pump^2 \sim \delta^{5/2}}},\nonumber
\end{align}
where we have introduced
\begin{equation}
    H_{mnq}(z) = 2\int_{-L/2}^{z} \exp\paren{\im\Delta k_{m,q-m}z'} I\conj_{nq}(z') \frac{\dif z'}{L}.
\end{equation}
Note that for the pump, we get no new terms at order $\mathcal O(\delta)$. It can also be seen that further Picard iterations do not yield any further corrections to either fields at $\mathcal O(\delta)$. Thus, we conclude that for small $\delta$, the crystal propagation implements a map ${\widetilde\alpha}{{}\sig}_m \mapsto {\widetilde\alpha}{{}\sig}'_m$ according to
\begin{subequations} \begin{align}
    {\widetilde\alpha}{{}\sig}'_m &= {\widetilde\alpha}{{}\sig}_m + \frac{\epsilon L}{\sqrt{T}} \sum_q {\widetilde\alpha}{{}\pump}_q {\widetilde\alpha}{{}\sig}\conj_{q-m} I_{mq}(L/2) \\
    &\qquad{} - \frac{\epsilon^2 L^2}{4T} \sum_{q,n} {\widetilde\alpha}{{}\sig}\conj_{q-m}{\widetilde\alpha}{{}\sig}_{n}{\widetilde\alpha}{{}\sig}_{q-n} H_{mnq}(L/2) \nonumber \\
    {\widetilde\alpha}{{}\pump}'_q &= {\widetilde\alpha}{{}\pump}_q - \frac12\frac{\epsilon L}{\sqrt{T}} \sum_m {\widetilde\alpha}{{}\sig}_m {\widetilde\alpha}{{}\sig}_{q-m} I\conj_{mq}(L/2),
\end{align} \end{subequations}
where the next-order correction to these equations are order $\mathcal O(\delta^{3/2})$.

Having completed the propagation through the crystal, we need to take care of the remaining cavity elements, which act only on the signal field. According to Sec.~\ref{sec:spopo}, this consists of applying two additional elements: a dispersion-compensating element to cancel the linear dispersion of the crystal, and an output coupler (or scattering loss) which attenuates the signal field.

First, in the short-crystal limit we are considering, the dispersion-compensation element can be modelled as a simple discrete transformation ${\widetilde\alpha}{{}\sig}_m \mapsto {\alpha\sig}_m$, i.e., \emph{physically} undoing the interaction frame transformation we performed in \eqref{eq:interaction-frame}. More precisely, the lab-frame coupled-wave equations \eqref{eq:coupled-wave-equations} involve dispersion and nonlinearity both acting simultaneously, and these differential operators do not commute in general. However, in the short-crystal limit, we can approximate the action of these two terms via a Suzuki-Trotter expansion as is done, for example, in the derivation of Fourier split-step methods for numerical pulse propagation. At leading order in $\delta$, then, we can factor out the linear dispersion as occurring \emph{after} the nonlinearity, which is immediately canceled by a separate dispersion-compensating element placed in series with the crystal. The effective roundtrip dynamics of these two elements then become well-described by \eqref{eq:coupled-wave-equations-IF}, but with ${\widetilde\alpha}{{}\sig}_m \mapsto {\alpha\sig}_m$ and so on.

Finally, the linear loss can be modeled as a beamsplitter with small field-outcoupling ratio $\ell_m \sim \delta^{1/2}$ for each signal Fourier mode ${\alpha\sig}_m$:
\begin{subequations} \begin{align}
    {\alpha\sig}_m &\mapsto \sqrt{1-\ell_m^2}{\alpha\sig}_m \approx \paren{1-\frac12\ell_m^2} {\alpha\sig}_m \\
    {\alpha^{(m)}_\text{s,out}} &= \ell_m {\alpha\sig}_m,
\end{align} \end{subequations}
where ${\alpha^{(m)}_\text{s,out}}$ is the outcoupled Fourier mode. Again, consistent with the low-gain, low-loss limit, we assume $\ell_m \ll 1$ and only consider the effect of the beamsplitter only up to order $\mathcal O(\ell_m^2 \sim \delta)$.

Putting all three of these effects together, the roundtrip recurrence relationship for the internal signal Fourier modes can be written, up to order $\mathcal O(\delta)$, as
\begin{align}
    {\alpha'\sig}_m &\approx \paren{1-\frac{\ell_m^2}{2}}{\alpha\sig}_m + \frac{\epsilon L}{\sqrt{T}} \sum_q {\widetilde\alpha}{{}\pump}_q {\alpha\sig}\conj_{q-m} I_{mq}\Paren{{\textstyle \frac L 2}} \nonumber\\
    &\qquad{} - \frac{\epsilon^2 L^2}{4T} \sum_{q,n} {\alpha\sig}\conj_{q-m}{\alpha\sig}_{n}{\alpha\sig}_{q-n} H_{mnq}\Paren{{\textstyle \frac L 2}}.
\end{align}
Again, $\alpha\sig'$ and $\alpha\sig$ denote the signal field immediately before and immediately after completing one roundtrip. If we also define ${\alpha^{(q)}_\text{p,out}}$ to be the post-crystal pump field ${\widetilde\alpha}{{}\pump}_q(z=L/2)$ after linear dispersion compensation, then
\begin{align}
    \!\!{\alpha^{(q)}_\text{p,out}} \approx {\alpha\pump} - \frac12\frac{\epsilon L}{\sqrt{T}} \sum_m {\alpha\sig}_m {\alpha\sig}_{q-m} I\conj_{mq}\Paren{{\textstyle \frac L 2}}.
\end{align}

Inspecting the equations above, we see that we can get a form very similar to \eqref{eq:mean-field-eoms} if we simply consider the finite-difference ratio $(\alpha'\sig - \alpha\sig)/T$, which is given by
\begin{align}
    \label{eq:classical-eom}
    &\frac{{\alpha'\sig}_m - {\alpha\sig}_m}{T} = -\frac{\ell_m^2}{2T}{\alpha\sig}_m + \frac{\epsilon L}{T^{3/2}} \sum_q {\alpha\pump}_q {\alpha\sig}\conj_{q-m} I_{mq}\Paren{{\textstyle \frac L 2}} \nonumber\\
    &\qquad{} - \frac{\epsilon^2 L^2}{4T^2} \sum_{q,n} {\alpha\sig}\conj_{q-m}{\alpha\sig}_{n}{\alpha\sig}_{q-n} H_{mnq}\Paren{{\textstyle \frac L 2}}.
\end{align}
Comparing this against \eqref{eq:mean-field-eoms}, we see that if we make the correspondence $\vbrak{\hat s_m} \leftrightarrow {\alpha\sig}_m$ and assume $({\alpha'\sig}_m - {\alpha\sig}_m)/T \rightarrow \dif\vbrak{\hat s_m}/\dif t$ in the limit $T \rightarrow 0$, then we can make the correspondences
\begin{subequations}\begin{align}
    \kappa_m &\leftrightarrow \frac{\ell_m^2}{2T} \\
    2\alpha^{(q)} f^{(q)}_{m,q-m} &\leftrightarrow \frac{\epsilon L}{T^{3/2}}{\alpha\pump}_q I_{mq}\Paren{\textstyle\frac L 2} \\
    2\xi^{(q)}_{nm} &\leftrightarrow \paren{\frac{\epsilon L}{2T}}^2 H_{mnq}\Paren{\textstyle \frac L 2}.
\end{align} \end{subequations}
To process these correspondences further, it is easy to show that
\begin{equation}
    I_{mq}\Paren{\textstyle \frac L 2} = \sinc\paren{\textstyle \frac 1 2 \Delta k_{m,q-m} L},
\end{equation}
whereas, e.g., via integration by parts, we have
\begin{align}
    \label{eq:cl_nl_tensor}
    & H_{mnq}\Paren{\textstyle \frac L 2} = \sinc\paren{\textstyle \frac 1 2 \Delta k_{m,q-m} L}\sinc\paren{\textstyle \frac 1 2 \Delta k_{n,q-n} L} \nonumber\\
    &\quad{}+ 2\im \paren{\Delta k_{m, q-m}-\Delta k_{n, q-n}}^{-1} L^{-1} \\
    &\qquad\quad{}\times \Bigl[ \cos\paren{\textstyle\frac{1}{2}\Delta k_{n, q-n} L} \sinc\paren{\textstyle\frac{1}{2}\Delta k_{m, q-m} L} \Bigr. \nonumber \\
    &\qquad\qquad\quad{}- \Bigl. \cos\paren{\textstyle\frac{1}{2}\Delta k_{m, q-m} L} \sinc\paren{\textstyle\frac{1}{2}\Delta k_{n, q-n} L}  \Bigr] \nonumber
\end{align}
Thus, this nonlinear interaction tensor in the classical theory is similar in form to its counterpart $\xi^{(q)}_{mn}$ in the quantum theory, as shown in \eqref{eq:xi_qmm'_final}. Putting everything together, we have that the mean field of the quantum theory and the classical theory are equivalent if we also impose the correspondences:
\begin{subequations} \begin{align}
    \label{eq:g0_correspondence}
    g_0 &\leftrightarrow \paren{\frac{\epsilon L}{2T}}^2 \\
    \Phi_{mn} &\leftrightarrow \frac 1 2 \Delta k_{mn} L \\
    \alpha^{(q)} &\leftrightarrow \frac{1}{\sqrt T} {\alpha\pump}_q.
\end{align} \end{subequations}

\section{Simulation method} \label{app:methods}
Here, we briefly review the key results from input-output theory we utilize in the numerical simulations of Sec.~\ref{sec:simulations}. Most of the formulas in this section follow the presentation in Ref.~\cite{Wiseman2010}.

We obtain the unconditional evolution of the system density matrix $\rho(t)$ using the master equation in Lindblad form:
\begin{equation}
    \frac{\dif\hat\rho}{\dif t} = -\im \Sbrak{\hat H_\text{sys}, \hat \rho} + \sum_i \hat L_i \hat \rho \hat L_i\dagg -\frac 1 2 \sum_i \Set{\hat L_i\dagg \hat L_i, \hat \rho}, \label{eq:master-equation}
\end{equation}
where $\hat H_{\text{sys}}$ is the system Hamiltonian and the $\hat L_i$ enumerate all the Lindblad operators of the system.

To compute the steady-state density matrix $\hat \rho_\text{ss}$ satisfying $\dif\hat \rho_\text{ss}/\dif t = 0$, we simulate \eqref{eq:master-equation} to a sufficiently large time $T$ such that $\hat \rho(t > T) \approx \hat \rho(T)$.

To obtain the steady-state squeezing spectrum that results from performing homodyne detection on a port represented by a particular Lindblad operator $\hat L$, we first compute the steady-state homodyne correlation function
\begin{equation}
    F_{\text{hom}}(\tau) = \tr\sbrak{\Paren{\hat L + \hat L\dagg}\hat A(\tau)} + \delta(\tau),
\end{equation}
where $A(\tau)$ is the solution to the differential equation
\begin{equation}
    \frac{\dif \hat A}{\dif \tau} = -\im \Sbrak{\hat H_\text{sys}, \hat A} + \sum_i \hat L_i \hat A \hat L_i\dagg - \frac 1 2 \sum_i\Set{\hat L_i\dagg\hat L_i, \hat A}
\end{equation}
with initial condition $\hat A(0) = \hat L \hat \rho_{\text{ss}} + \hat \rho_{\text{ss}} \hat L\dagg$. The squeezing spectrum is then the Fourier transform of the correlation function:
\begin{equation}
\label{eq:S_hom}
  S_\text{hom}(\omega) \coloneqq 2\mathrm{Re}\paren{\int_{0}^\infty e^{-\im \omega \tau} F_{\text{hom}}(\tau) \, \dif\tau}.
\end{equation}
As described in Sec.\ref{sec:simulations}, the squeezing spectrum is computed at the optimal squeezing angle. Thus, the Lindblad operator that we use in these simulations is given by $\hat L = \sqrt{2\kappa}e^{-\im \theta_\text{opt}}{\hat{S}}_1$.

To obtain conditional evolution, we simulate the stochastic Schr\"{o}dinger equation (SSE), by solving the unnormalized SSE numerically and normalizing the state at every timestep. The unnormalized SSE is
\begin{equation}
    \dif\ket{\overline\psi} = \Bigl(\hat f \, \dif t + \sum_i \hat g_i \, \dif W_i\Bigr)\ket{\overline\psi},
\end{equation}
where $\dif W_i$ are differentials of independent standard Weiner processes, and the deterministic and stochastic components of the stochastic differential equation are, respectively,
\begin{subequations} \begin{align}
    \hat f &\coloneqq -\im \hat H_{\text{sys}} + \sum_i \Bigl(-\frac 1 2 \hat L_i\dagg \hat L_i  + \Vbrak{\hat L_i + \hat L_i\dagg} \hat L_i\Bigr) \\
    \hat g_i &\coloneqq \hat L_i .
\end{align} \end{subequations}

\providecommand{\noopsort}[1]{}\providecommand{\singleletter}[1]{#1}%

\end{document}